\documentclass[12pt,a4paper]{article}
\usepackage{jheppub}  
\usepackage{amssymb} 
\usepackage{amsmath}
\usepackage{mathtools}
\usepackage{amsfonts}    
\usepackage{dsfont}
\usepackage{pdfpages}
\usepackage{verbatim}
\usepackage{tensor}
\usepackage{mathrsfs}
\usepackage{textgreek} 
\usepackage[mathscr]{euscript}
\usepackage{tikz}
\usetikzlibrary{decorations.pathreplacing, decorations.markings,calc,shapes.misc,decorations.pathmorphing,patterns.meta, math}

\newcommand{\ba}{\begin{align}}

\newcommand{\be}{\begin{equation}}
\newcommand{\ee}{\end{equation}}
\def\bd{\begin{tikzpicture}}
\def\ed{\end{tikzpicture}}

\DeclareMathOperator\tr{tr}
\DeclareMathOperator\Tr{Tr}
\newcommand\e{\text{e}}

\allowdisplaybreaks[1]

\newcommand\PSL{\text{PSL}}

\newcommand\U{\text{U}}

\newcommand\Diff{\text{Diff}}

\newcommand\ZZ{\mathbb{Z}}
\newcommand\RR{\mathbb{R}}

\newcommand\LL{\mathbb{L}}

\let\H\relax
\DeclareMathOperator\H{H}

\newcommand{\bM}{\overline{\mathcal{M}}}
\newcommand{\bC}{\overline{\mathcal{C}}}
\newcommand{\M}{\mathcal{M}}
\renewcommand\d{{\mathrm d}}
\renewcommand\i{{\mathrm i}}

\title{2D dilaton gravity and the Weil-Petersson volumes with conical defects}

\author[1]{Lorenz Eberhardt}\emailAdd{elorenz@ias.edu}
\author[1,2]{\!\!, Gustavo J.\ Turiaci}\emailAdd{turiaci@ias.edu}
\affiliation[1]{Institute for Advanced Study, Einstein Drive, Princeton, NJ, USA}
\affiliation[2]{Physics Department, University of Washington, Seattle, WA, USA}

\abstract{We derive the Weil-Petersson measure on the moduli space of hyperbolic surfaces with defects of arbitrary opening angles and use this to compute its volume.
We conjecture a matrix integral computing the corresponding volumes and confirm agreement in simple cases. We combine this mathematical result with the equivariant localization approach to Jackiw-Teitelboim gravity to justify a proposed exact solution of pure 2d dilaton gravity for a large class of dilaton potentials.}

\begin{document}

\maketitle

\makeatletter
\g@addto@macro\bfseries{\boldmath}
\makeatother


\section{Introduction}
Theories of two dimensional dilaton gravity coupled to matter appear naturally when describing the low energy dynamics of near extremal black holes in higher dimensions. These two dimensional theories have been a very fruitful theoretical laboratory to explore the structure and general properties of the gravitational path integral. Due to its solvability, an emphasis has been put in Jackiw-Teitelboim (JT) gravity which consists of pure gravity with a linear dilaton potential, see \cite{Mertens:2022irh} for a recent review. 

To construct models describing higher dimensional black holes, one has to move away from pure gravity and add propagating matter degrees of freedom. This is a very difficult problem and some progress was made in \cite{Jafferis:2022wez}. Instead, in this paper we address a different issue: in realistic models described by 2d dilaton gravity the dilaton potential is not linear. It would be interesting therefore to have a solvable model of 2d pure dilaton gravity with a generic dilaton potential, that one could eventually couple to matter.

Leveraging on the techniques developed to solve JT gravity, \cite{Maxfield:2020ale} and \cite{Witten:2020wvy} proposed to model a modification of the dilaton potential by deforming JT gravity by a gas of defects. The weight of the defect in the path integral, and the deficit angle, appear as parameters of the new dilaton potential. In those references the theory was solved in a regime involving sharp defects, with deficit angles bigger than $\pi$. This restriction does not allow to study a semiclassical limit, where one could hope to see the sum over gas of defects turn into a smooth geometry. This requires considering blunt defects with deficit angles close to $0$. 

In \cite{Turiaci:2020fjj} a conjecture was made for what the gravitational path integral should be when deforming JT by a gas of defects with arbitrary angles. The result was motivated by taking a limit of the minimal string theory, but no prescription about how to compute it from first principles was given. The purpose of this paper is to fill this gap. On the way, we also clarify some connections between Weil-Petersson volumes of moduli space of hyperbolic surfaces with conical defects, and the minimal string.

The paper has two parts. In the first part, presented in section \ref{sec:Weil Petersson form}, we prove new mathematical results regarding the Weil-Petersson measure on the moduli space of hyperbolic surfaces with conical deficits of arbitrary angles from $0$ to $2\pi$. In the second part, section \ref{sec:JTRMT}, we apply this to elucidate the meaning of the solution proposed by \cite{Turiaci:2020fjj}. In the rest of this introduction we summarize the results in some detail.

\subsection*{Mathematical results: Weil-Petersson volumes with conical defects}
The moduli space of hyperbolic surfaces of genus $g$ with $n$ geodesic boundaries and its volume have been intensely studied in both the math and physics literature, see e.g.\ \cite{Wolpert:1985,Mirzakhani:2006fta, Mirzakhani:2006eta, Mirzakhani:2010pla, Saad:2019lba}. From a physics point of view, it mainly appeared in the path-integral formulation of JT gravity for which the partition function is given by the Weil-Petersson volume of surfaces with the given topology.

JT gravity appears as the large $p$ limit of $(2,2p+1)$ minimal string theory when considered as a 2d theory of gravity \cite{Seiberg_largep}, see also \cite[Appendix F]{Mertens:2020hbs}. In the minimal string, it is natural to also consider vertex operators. They are labelled by primary fields of the $(2,2p+1)$ Virasoro minimal model, which we can take to run from $0 \le k \le p-1$ with $k \in \ZZ$. The interpretation of these vertex operators in a large $p$ limit is that they create a conical defect in the worldsheet and hence the minimal string correlation functions in a large $p$ limit become the Weil-Petersson volume of the moduli space of hyperbolic surfaces with conical defects.

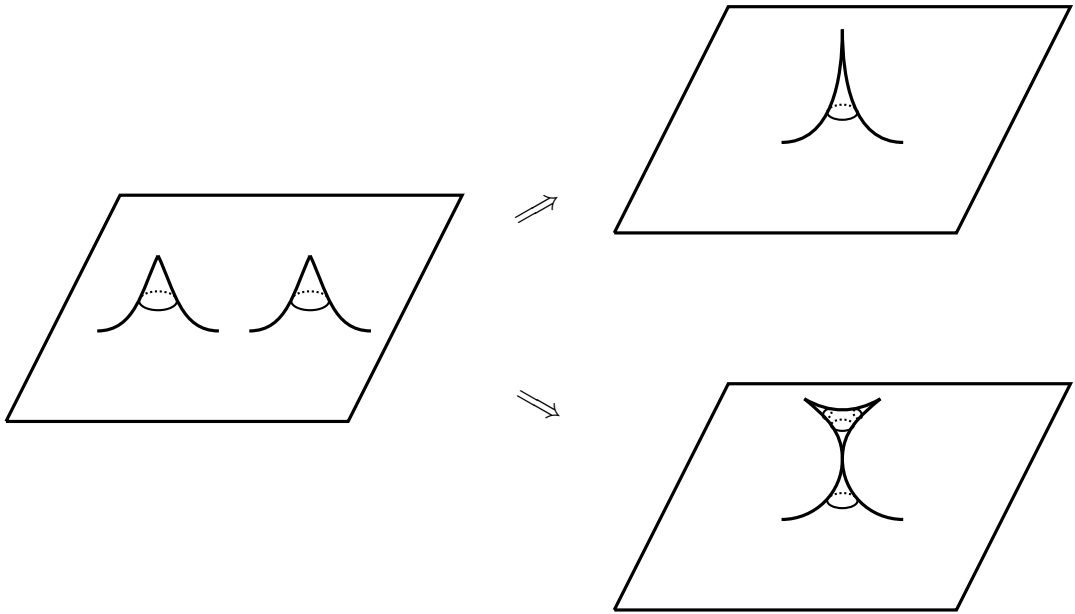
\begin{figure}
    \centering
    \begin{tikzpicture}
        \begin{scope}
            \draw[very thick] (-3,-1.5) to (1.5,-1.5) to (3,1.5) to (-1.5,1.5) to (-3,-1.5);
            \draw[very thick, out=0, in=245] (-1.8,-.3) to (-1,.7);
            \draw[very thick, out=180, in=295] (-.2,-.3) to (-1,.7);
            \draw[very thick, out=0, in=245] (.2,-.3) to (1,.7);
            \draw[very thick, out=180, in=295] (1.8,-.3) to (1,.7);
            \begin{scope}[shift={(-1,.1)}, yscale=.5]
                \draw[thick] (-.25,0) arc (180:360:.25);
                \draw[thick, densely dotted] (.25,0) arc (0:180:.25);
            \end{scope}
            \begin{scope}[shift={(1,.1)}, yscale=.5]
                \draw[thick] (-.25,0) arc (180:360:.25);
                \draw[thick, densely dotted] (.25,0) arc (0:180:.25);
            \end{scope}
        \end{scope}
        
        \begin{scope}[shift={(8,2.5)}]
            \draw[very thick] (-3,-1.5) to (1.5,-1.5) to (3,1.5) to (-1.5,1.5) to (-3,-1.5);
            \draw[very thick, out=0, in=270] (-.8,-.3) to (0,1.2);
            \draw[very thick, out=180, in=270] (.8,-.3) to (0,1.2);
            \begin{scope}[shift={(0,.1)}, yscale=.5]
                \draw[thick] (-.2,0) arc (180:360:.2);
                \draw[thick, densely dotted] (.2,0) arc (0:180:.2);
            \end{scope}
            \node[rotate=30] at (-4,-1.2) {$\Longrightarrow$};
        \end{scope}
        
        \begin{scope}[shift={(8,-2.5)}]
            \draw[very thick] (-3,-1.5) to (1.5,-1.5) to (3,1.5) to (-1.5,1.5) to (-3,-1.5);
            \draw[very thick, out=0, in=270] (-.8,-.3) to (0,.5);
            \draw[very thick, out=90, in=220] (0,.5) to (.5,1.3);
            \draw[very thick, out=180, in=270] (.8,-.3) to (0,.5);
            \draw[very thick, out=90, in=320] (0,.5) to (-.5,1.3);
            \draw[very thick, bend right=30] (-.5,1.3) to (.5,1.3);
            \begin{scope}[shift={(0,-.05)}, yscale=.5]
                \draw[thick] (-.2,0) arc (180:360:.2);
                \draw[thick, densely dotted] (.2,0) arc (0:180:.2);
            \end{scope}
            \begin{scope}[shift={(0,.95)}, yscale=.5]
                \draw[thick] (-.15,0) arc (180:360:.15);
                \draw[thick, densely dotted] (.15,0) arc (0:180:.15);
            \end{scope}
            \begin{scope}[shift={(.2,1.1)}]
                \node[rotate=-55] at (0,0) {%
                    \begin{tikzpicture}
                        \begin{scope}[yscale=.7]
                        \draw[thick, densely dotted] (-.07,0) arc (180:360:.07);
                        \draw[thick] (.07,0) arc (0:180:.07);
                        \end{scope}
                    \end{tikzpicture}
                };
            \end{scope}
            \begin{scope}[shift={(-.2,1.1)}]
                \node[rotate=55] at (0,0) {%
                    \begin{tikzpicture}
                        \begin{scope}[yscale=.7]
                        \draw[thick, densely dotted] (-.07,0) arc (180:360:.07);
                        \draw[thick] (.07,0) arc (0:180:.07);
                        \end{scope}
                    \end{tikzpicture}
                };
            \end{scope}
            \node[rotate=-30] at (-4,1.2) {$\Longrightarrow$};
        \end{scope}
        
    \end{tikzpicture}
    \caption{The two behaviors of two merging conical singularities.}
    \label{fig:defect merging}
\end{figure}

Computing the volume of the moduli space of cone surfaces is a surprisingly subtle problem and has only partially been solved in the literature \cite{Tan:2006, Do_cone, Witten:2020wvy}. The trouble is that there are various `phase transitions' and the volume is in fact not a smooth function of the conical defect angles. To see geometrically what is going on consider Figure~\ref{fig:defect merging}.
It shows a portion of the surface with two conical defects. One of the moduli of the surface is the relative position of the two defects. Hence we can consider what happens when the two defects approach each other. Depending on the precise choice of the defect angle, there are two different scenarios what happens geometrically. If the defects are sufficiently `blunt', then they can merge to one conical defect angle as they collide. The new defect angle is now the sum of the original two defect angles. Since the defect angle can be at most $2\pi$, the two defects can only merge as long as the sum of their angles is less than $2\pi$.
If the sum exceeds $2\pi$, the situation is analogous to the behavior for geodesic boundaries. The two defects can actually never merge. Instead, there is a geodesic that encircles the two defects and the geodesic pinches in the appropriate degeneration. As we get closer to the degeneration, the two original defects protrude further and further from the rest of the surface. In the limit, they form a three-punctured sphere together with the node where the geodesic pinched. The surface hence splits into two components that are connected at the single node. This is the second behavior sketched in 
Figure~\ref{fig:defect merging}. We call such a pair of defects sharp. The same two scenarios can appear in the more general case where $n$ conical defects approach each other.

Weil-Petersson volumes were computed by Mirzakhani using a geometric recursion relation \cite{Mirzakhani:2006fta}. A crucial ingredient of this recursion relation is the existence of geodesics that divide the surface into smaller pieces. For blunt defects such geodesics do not exist and the logic of the recursion relation breaks down. 

We propose a simple solution to compute these volumes in generality in this paper by following a different route. We make use of the well-developed intersection theory on moduli space. These techniques were applied to JT gravity in \cite{Okuyama:2019xbv,Okuyama:2020ncd}. 
It is a useful fact that the moduli space of cone surfaces $\M_{g;\, \alpha_1,\dots,\alpha_n}$ can be given a natural complex structure for closed surfaces and hence one may invoke the power of algebraic geometry. 
For the case with geodesic boundaries or cusps, the moduli space of hyperbolic surfaces is isomorphic to the Deligne-Mumford compactification $\bM_{g,n}$ of moduli space. For conical defects, this is not quite true. As we already discussed we allow points to coincide as long as their defect angles are blunt. Hence the moduli space $\bM_{g;\, \alpha_1,\dots,\alpha_n}$ might correspond to a smaller compactification. The above discussion implies that the moduli space of cone surface $\bM_{g;\, \alpha_1,\dots,\alpha_n}$ changes actually discontinuously with the defect angles $\alpha_i$. In more technical terms, $\bM_{g,n}$ is a blow up of $\bM_{g;\alpha_1,\dots,\alpha_n}$. In particular, there is a map $\bM_{g,n} \longrightarrow \bM_{g;\alpha_1,\dots,\alpha_n}$ that forgets the blow up, i.e.\ $\bM_{g,n}$ is a redundant parametrization of cone surfaces. It is thus convenient to always work with $\bM_{g,n}$.

The volume of $\bM_{g;\, \alpha_1,\dots,\alpha_n}$ can then be found by computing
\be 
V_{g,n}(\alpha_1,\dots,\alpha_n)=\int_{\bM_{g;\, \alpha_1,\dots,\alpha_n}} \e^{\omega_\text{WP}}\ ,
\ee
where $\omega_\text{WP}$ is the Weil-Petersson form. The moduli space has complex dimension $3g-3+n$ and only that term in the power series expansion of the exponential function contributes. We will denote the Weil-Petersson form on $\bM_{g,n}$ also by $\omega_\text{WP}$ and the same formula for the volume also holds on $\bM_{g,n}$. The task of computing the volume then boils down to identify the correct cohomology class $[\omega_\text{WP}] \in \H^2(\bM_{g,n},\RR)$ that represents the Weil-Petersson form on moduli space.\footnote{All cohomology in this paper will be taken over the reals and we will hence surpress it in the following.} After that, one can use standard intersection theory on $\bM_{g,n}$ to compute the volumes. The main non-trivial ingredient in the computation of intersection numbers of $\bM_{g,n}$ is Witten's conjecture/Kontsevich's theorem \cite{Witten:1990hr, Kontsevich:1992ti} to which all other intersection numbers can be reduced in a rather simple fashion. An introductory account of this is given for example in \cite{Zvonkine_intro}. 

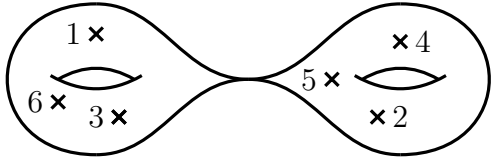
\begin{figure}
    \centering
    \begin{tikzpicture}
        \draw[very thick, out=180, in=180, looseness=2] (-2,1) to (-2,-1);
        \draw[very thick, out=0, in=180] (-2,1) to (0,0);
        \draw[very thick, out=0, in=180] (-2,-1) to (0,0);
        \draw[very thick, out=0, in=180] (0,0) to (2,1);
        \draw[very thick, out=0, in=180] (0,0) to (2,-1);
        \draw[very thick, out=0, in=0, looseness=2] (2,1) to (2,-1);
        \draw[very thick, bend right=30] (-2.6,.05) to (-1.4,.05); 
        \draw[very thick, bend left=30] (-2.5,0) to (-1.5,0); 
        \draw[very thick, bend right=30] (1.4,.05) to (2.6,.05); 
        \draw[very thick, bend left=30] (1.5,0) to (2.5,0); 
        \draw  (-2,.6) node[cross out, draw=black, very thick, minimum size=5pt, inner sep=0pt, outer sep=0pt] {};
        \node at (-2.3,.6) {1};
        \draw  (-1.7,-.5) node[cross out, draw=black, very thick, minimum size=5pt, inner sep=0pt, outer sep=0pt] {};
        \node at (-2,-.5) {3};
        \draw  (-1.7,-.5) node[cross out, draw=black, very thick, minimum size=5pt, inner sep=0pt, outer sep=0pt] {};
        \node at (-2.8,-.3) {6};
        \draw  (-2.5,-.3) node[cross out, draw=black, very thick, minimum size=5pt, inner sep=0pt, outer sep=0pt] {};
        \node at (2,-.5) {2};
        \draw  (1.7,-.5) node[cross out, draw=black, very thick, minimum size=5pt, inner sep=0pt, outer sep=0pt] {};
        \node at (2.3,.5) {4};
        \draw  (2,.5) node[cross out, draw=black, very thick, minimum size=5pt, inner sep=0pt, outer sep=0pt] {};
        \node at (.8,0) {5};
        \draw  (1.1,0) node[cross out, draw=black, very thick, minimum size=5pt, inner sep=0pt, outer sep=0pt] {};
    \end{tikzpicture}
    \caption{The boundary divisor for the class $\delta_{1,\{1,3,6\}}=\delta_{1,\{2,4,5\}} \in \H^2(\bM_{2,6})$.}
    \label{fig:separating divisor}
\end{figure}

To explain the formula that we derive in this paper, we now recall some basic constructions in $\bM_{g,n}$. There are $n$ natural holomorphic line bundles $\LL_1,\dots,\LL_n$ over $\bM_{g,n}$ whose first Chern classes are traditionally denoted by $\psi_1,\dots,\psi_n$ in the literature. Somewhat informally, $\LL_i$ is the line bundle whose fiber over the punctured surface $\Sigma$ consists of the cotangent space at the $i$-th marked point $z_i \in \Sigma$, i.e. $\LL_i\big|_{\Sigma}=T_{z_i}^*\Sigma$.\footnote{This definition is not adequate to describe the the behavior of sections near degenerations of moduli space. A better definition is to define $\LL_i$ in terms of the relative dualizing sheaf of the universal curve $\bC_{g,n}$. We will not need to work with the actual line bundles $\LL_i$ and will thus not get into the details of these subtleties.}
There is also a natural class that is usually called $\kappa_1$ in the literature. $2\pi^2 \kappa_1$ is known to represent the Weil-Petersson form on moduli space in the case in which all the marked points are cusps \cite{Wolpert:1983,  Wolpert:1986}. A more algebraic definition is as follows. Letting $\pi:\bM_{g,n+1} \longrightarrow \bM_{g,n}$ be the forgetful morphism that forgets the last marked point, we can define $\kappa_1=\pi_*(\psi_{n+1}^2)$. Here we recall that pushforward in cohomology means fiberwise integration. Since the fiber of the map $\pi$ is complex one-dimensional and $\psi_{n+1}^2$ is a 4-form, $\kappa_1$ is hence indeed a 2-form.\footnote{To explain the notation, it is useful to note that there is a natural generalization $\kappa_m=\pi_*(\psi_{n+1}^{m+1})$, which are the so-called Mumford-Morita-Muller classes. We will however only need $\kappa_1$ in this paper.}
Finally, further natural classes in $\H^2(\bM_{g,n})$ are provided by the Poincar\'e duals of boundary divisors. Such classes can informally be written by forms (or more precisely currents) that have delta-function support on the boundary divisor. In physics language, such a form will hence correspond to a contact term on the worldsheet that appears when two or more vertex operators collide or when the surface degenerates in any other way. We denote these classes by $\delta_{h,I} \in \H^2(\bM_{g,n})$. Here $I \subset \{1,\dots,n\}$ and $0 \le h \le g$. This class is the Poincar\'e dual of the degeneration depicted in Figure~\ref{fig:separating divisor}, where the surface splits into two parts of genus $h$ and $g-h$ and the first part contains the marked points $I$, while the second part contains the remaining marked points $I^c$. Obviously we have by construction $\delta_{h,I}=\delta_{g-h,I^c}$. We also have the requirement $|I| \ge 2$ for $h=0$ or $|I^c|\ge 2$ for $h=g$, since there is no hyerbolic metric on a sphere with one or two punctures. There is also a boundary class $\delta_\text{irr}$ that corresponds to the non-separating divisor in $\bM_{g,n}$. It will however not make an appearance in this paper.

With these preparations, we can now state the main formula that we derive in this paper. The class  $[\omega_\text{WP}]$ for the moduli space $\bM_{g; \alpha_1,\dots,\alpha_m;\, b_1,\dots,b_{n-m}}$ of hyperbolic surfaces with $m$ conical defects and $n-m$ geodesic boundaries is in general given by
\be 
\frac{[\omega_\text{WP}]}{2\pi^2}=\kappa_1-\sum_{i=1}^m \alpha_i^2 \psi_i+\frac{1}{4\pi^2} \sum_{i=m+1}^n b_i^2 \psi_i +\hspace{-.5cm}\sum_{\begin{subarray}{c} I \subset \{1,\dots,m\} \\ 1-\sum_{i \in I} (1-\alpha_i) \ge 0 \end{subarray}} \hspace{-.5cm}\bigg(1-\sum_{i \in I} (1-\alpha_i) \bigg)^2 \delta_{0,I}\ . \label{eq:Weil-Petersson form with boundary classes}
\ee
Here we assumed that the first $m$ marked points are conical defects with deficit angles $2\pi(1-\alpha_i)$ and the marked points $m+1,\dots,n$ are geodesic boundaries with boundary lengths $b_i$. Notice that the condition $1-\sum_{i \in I}(1-\alpha_i) \ge 0$ precisely corresponds to the condition that the defects labelled by the subset $I$ can merge, i.e.\ realize the first scenario in Figure~\ref{fig:defect merging}. The first three terms in this formula are the `naive answer' that has appeared before in the literature \cite{Witten:2020wvy}. It is obtained by noticing that a sharp defect behaves essentially like a geodesic boundary whose length is formally imaginary. One can then use the known formula for the case of only geodesic boundaries derived by Mirzakhani \cite{Mirzakhani:2006fta}. The last term should hence be understood as a correction term that is only present when at least some pair of defects is blunt. In physics language it represents a contact term on the worldsheet.

We will show that \eqref{eq:Weil-Petersson form with boundary classes} is uniquely fixed by requiring natural properties that we will explain in detail in Section~\ref{sec:Weil Petersson form}. We also show that the volumes defined in this way satisfy analogues of the string and dilaton equation. They deal with the behavior of the volumes when one of the defect angles approaches 0 and hence the surface becomes completely regular at the puncture. These two identities read
\begin{subequations} \label{eq:string and dilaton equations}
\begin{align}
    V_{g,m+1,n-m}(\boldsymbol{\alpha},\alpha_{m+1};\boldsymbol{b}) \Big|_{\alpha_{m+1}=1}&=\sum_{i=1}^{n-m} \int \d  b_i\ b_i\,  V_{g,m,n-m}(\boldsymbol{\alpha};\boldsymbol{b})\ , \label{eq:string equation}\\
    \frac{\d V_{g,m+1,n-m}(\boldsymbol{\alpha},\alpha_{m+1};\boldsymbol{b})}{\d\alpha_{m+1}} \Big|_{\alpha_{m+1}=1}& =2\,\chi(\Sigma_{g,\boldsymbol{\alpha};\boldsymbol{b}}) V_{g,m,n-m}(\boldsymbol{\alpha};\boldsymbol{b})\ . \label{eq:dilaton equation}
\end{align}
\end{subequations}
The integral on the right hand side is a shortcut for $\int_0^{b_i} \d b_i'$. Here,
\be 
\chi(\Sigma_{g,\boldsymbol{\alpha};\boldsymbol{b}})=2-2g-\sum_{i=1}^m(1-\alpha_i)-(n-m)
\ee
is the natural generalization of the Euler characteristic to the case of cone surfaces. In particular, it is the quantity that the integral over the curvature computes in this case via the Gauss-Bonnet theorem. Since the curvature is constant negative, $-\chi(\Sigma_{g,\boldsymbol{\alpha};\boldsymbol{b}})$ is the area of the cone surface.
For the case of only one conical defect, these identities were derived in \cite{Do_cone}.

\subsection*{Application to 2d dilaton gravity}

As explained above, we defined a measure to compute volumes on the moduli space of hyperbolic surfaces with cone points of arbitrary angles. We apply this to the problem of solving 2d pure dilaton gravity with an arbitrary dilaton potential. These are theories involving a two dimensional metric $g_{\mu\nu}$ on a surface $\mathcal{M}$ and dilaton $\phi$ with an action in Euclidean signature given by
\begin{equation}
    I[g,\phi] = - S_0 \chi - \frac{1}{2} \int_{\mathcal{M}} \sqrt{g} (\phi R+U(\phi)) - \oint_{\partial \mathcal{M}}  \sqrt{h} \phi (K-1).
\end{equation}
The first term is proportional to the Euler characteristic of the space, $S_0$ is a parameter of the theory, and $U(\phi)$ is an a priori arbitrary function of the dilaton $\phi$ called the dilaton potential. The last term includes the Gibbons-Hawking-York piece, to make the variational problem well-defined for Dirichlet boundary conditions on the metric, and a holographic counterterm to make the action finite on the hyperbolic disk. A connected geometry with a number of handles $g$ and $n$ boundaries is suppressed by a factor of $(\e^{-S_0})^{2g+n-2}$, and therefore the large $S_0$ limit organizes the path integral into sectors of fixed topology.

Let us recapitulate what we know about 2d dilaton gravity. First, it was proven by Saad, Shenker and Stanford (SSS) that the gravitational path integral of pure JT gravity is dual to a random matrix integral with a specific spectral curve \cite{Saad:2019lba}, derived from a leading order in $S_0$ density of states $\e^{S_0} \rho_{\sf JT}(E)$ with
\be \label{eq:introrhojt}
    \rho_{\sf JT}(E)= \frac{\sinh(2\pi \sqrt{E})}{4\pi^2}\ , \qquad E>0\ .
\ee
We denote by $E$ the eigenvalues of the matrix, since they correspond to energy eigenvalues in the holographic dual interpretation. This corresponds to a specific choice of the dilaton potential
\begin{equation}
    U_{\sf JT}(\phi) = 2\phi\ .
\end{equation}
In this case the path integral over the dilaton localizes the gravity path integral into geometries with constant negative curvature, which are asymptotically nearly-AdS${}_2$ \cite{Maldacena:2016upp, Jensen:2016pah, Engelsoy:2016xyb}. To prove this, SSS evaluates the partition function with a fixed number of boundaries and handles and relates it to the Weil-Petersson volumes of smooth hyperbolic surfaces. This relies crucially on the existence of geodesics separating the asymptotically AdS${}_2$ boundaries from the handles present in the interior of the geometry. The remain of the proof then follows from the known connection between these volumes and random matrices \cite{Mirzakhani:2006fta,Eynard:2007fi}. 

In \cite{Maxfield:2020ale} and \cite{Witten:2020wvy} this duality was extended to deformations of JT gravity, where one performs the gravity path integral including a gas of sharp defects with a deficit angle between $\pi$ and $2\pi$. Given a fixed number of boundaries, the partition function is a double expansion in the number of handles and the number of defects. It is argued that adding a single defect species is equivalent to shifting the dilaton potential by 
\begin{equation}
U(\phi) \to U(\phi) + \lambda \e^{-2\pi(1-\alpha) \phi}.
\end{equation}
The proof of this duality relies on the fact that the WP volumes with sharp defects are simply related to the original WP volumes with geodesic boundaries Mirzakhani computed \cite{Tan:2006, Do_cone}. 

At each order in $\lambda$, the geometry in the bulk is singular due to the presence of defects. It is a non-trivial expectation that the resummation over the gas of defects generates a smooth new geometry, not necessarily hyperbolic. This analysis is only relevant when quantum corrections are small. Looking at \eqref{eq:introrhojt} for pure JT gravity, one sees that quantum corrections are small at large energies $E\gg1$. The high energy part of the spectrum can be associated to regions in the bulk with large dilaton. If $\alpha<1$ we see that the modification of the dilaton potential does not affect the high energy sector. To test whether the sum over defects really gives rise to a new geometry we need to take the limit of small $1-\alpha$, and therefore consider blunt defects. 

A proposal was made in \cite{Turiaci:2020fjj} that relates the path integral of JT gravity deformed by a gas of generic defects, with arbitrary deficit angles from $0$ to $2\pi$, with a matrix integral. The integral is over hermitian matrices with a matrix potential such that, after a double scaling limit, the leading order density of state is $\e^{S_0} \rho_{\sf dJT}(E)$ with
\be \label{eq:DOSGDJTintro}
    \rho_{\sf dJT}(E)= \frac{1}{2\pi} \int_{\mathcal{C}} \frac{\d y}{2\pi {\rm i}} \e^{2\pi y} \tanh^{-1} \left(\sqrt{\frac{E-E_0}{y^2-2W(y)-E_0}}\right)\ , \qquad E>E_0\ .
\ee
 The dependence on the deficit parameters appears through the function $W(y) = \sum_i \lambda_i \e^{-2\pi (1-\alpha_i )y}$. The value $E_0$ is found by demanding the density of states vanishes at $E_0$ with a square-root edge, and implicitly depends on $W(y)$ in a complicated way. The contour $\mathcal{C}$ is along the imaginary $y$-axis to the right of all singularities. This coincides with the solution of \cite{Maxfield:2020ale,Witten:2020wvy} when all defects are sharp. This proposal was motivated by a conjectural connection between JT gravity and the minimal string theory, but the analysis of \cite{Turiaci:2020fjj} gave no clue of how to perform the gravity calculation. 
 
 Using the mathematical results described before, we show in some examples how the gravitational path integral of theories with blunt defects matches with the matrix model derived from \eqref{eq:DOSGDJTintro}. More concretely, the genus expansion  of a matrix model with density of states \eqref{eq:DOSGDJTintro} is a generating function when expanded in $\lambda$ of the WP volumes at fixed genus and number of defects, introduced in section \ref{sec:Weil Petersson form}. We should study two cases separately. First, consider the path integral over surfaces with handles and defects such that there are geodesics separating the structure in the interior to the AdS${}_2$ boundaries. In this case we propose that the result in \cite{Turiaci:2020fjj} computes precisely the Weil-Petersson volumes defined in section \ref{sec:Weil Petersson form}. Second, there are some simple cases where there are no geodesics. This happens when we have a small number of defects inside the hyperbolic disk. We argue how the path integral on such spaces can be computed using the equivariant localization approach of \cite{Eberhardt:2022wlc}, and match it with the results from the matrix integral of \cite{Turiaci:2020fjj}. These two cases exhaust all the possibilities.

Proving the duality between the matrix integral with spectral curve derived from \eqref{eq:DOSGDJTintro} and the Weil-Petersson volumes with conical deficits derived from \eqref{eq:Weil-Petersson form with boundary classes} is an open mathematical problem. The spectral curve is complicated enough that a simple extension of the methods in \cite{Eynard:2007fi} does not seem viable.

\section{The Weil-Petersson form on the moduli space of cone surfaces} \label{sec:Weil Petersson form}

Mirzakhani showed in  \cite{Mirzakhani:2006eta} that the cohomology class of the Weil-Petersson form on the moduli space of surfaces with $n$ geodesic boundaries of length $b_1,\dots,b_n$ is given by
\be 
[\omega_\text{WP}]=2\pi^2 \kappa_1+\frac{1}{2}\sum_i b_i^2 \psi_i\ . \label{eq:Weil Petersson form geodesic boundaries}
\ee
The definition of the $\kappa$- and $\psi$-classes was explained in the Introduction, see also \cite{Zvonkine_intro}. The special case of $b_i=0$, i.e.\ a surface with only cusp singularities is a result due to Wolpert \cite{Wolpert:1983,  Wolpert:1986}.

The Weil-Petersson form on moduli space descends from the Weil-Petersson form on Teichmuller space, which in turn can be identified with a component of the moduli space of flat $\PSL(2,\RR)$-bundles on the Riemann surface. Thus there is a simple gauge-theory formula for the Weil-Petersson form \cite{AtiyahBott, Goldman_symplectic}, 
\be 
\omega_\text{WP}(\delta A,\delta A')\equiv \frac{1}{4\pi} \int_\Sigma \tr \left(\delta A \wedge \delta A'\right)\ , \label{eq:Atiyah Bott symplectic form}
\ee
where $\delta A$ and $\delta A'$ are small perturbations of a connection on $\Sigma$ and hence represent tangent vectors on the space of \emph{all} $\PSL(2,\RR)$ connections. Via symplectic reduction this formula descends to a non-degenerate symplectic form on the space of flat $\PSL(2,\RR)$ connections up to gauge redundancy. From this point of view, one specifies the monodromy of the gauge field around the punctures of the surface. 
Geodesic boundaries correspond to a monodromy in hyperbolic conjugacy classes of $\PSL(2,\RR)$, while conical defects correspond to monodromies in  elliptic conjugacy classes. More precisely, the monodromy matrices in the two cases take the form\footnote{Strictly speaking these monodromies should be interpreted as monodromies of the universal cover of $\PSL(2,\RR)$ so that e.g.\ the monodromy with $\alpha=0$ and $\alpha=1$ becomes inequivalent.}
\be 
M \sim \begin{pmatrix}
\e^{\frac{b}{2}} & 0 \\ 0 & \e^{-\frac{b}{2}}
\end{pmatrix}\ , \qquad M \sim \begin{pmatrix}
\cos(\pi \alpha) & \sin(\pi \alpha) \\ -\sin(\pi \alpha) & \cos(\pi \alpha)
\end{pmatrix}\ .
\ee
The parameter $\alpha\in [0,1]$ characterizes the conical defect as in the Introduction. The defect angle is related to $\alpha$ as $\theta=2\pi(1-\alpha)$. Hence $\alpha=0$ corresponds to a cusp and $\alpha=1$ corresponds to no defect at all. Since the eigenvalues of the two matrices are related by
\be 
b \sim 2\pi \i \alpha\ ,
\ee
one concludes that a geodesic boundary with imaginary length is formally equivalent to a conical defect, at least for the purpose of computing volumes of moduli spaces. The trace in \eqref{eq:Atiyah Bott symplectic form} is invariant under conjugations, including complex ones. One may hence be tempted to think that \eqref{eq:Weil Petersson form geodesic boundaries} implies that
\be 
\frac{[\omega^\text{WP}]}{2\pi^2}\overset{?}{=}\kappa_1-\sum_i \alpha_i^2 \psi_i \label{eq:wrong WP form defect}
\ee
for the cohomology class of the Weil-Petersson form for a surface with $n$ conical defects.

This identification is in general incorrect. To see a simple example where this fails, consider a four-punctured sphere. The generalized Gauss-Bonnet theorem implies that the four-punctured sphere carries a hyperbolic metric under condition that
\be 
\sum_{i=1}^4 \alpha_i<2\ . \label{eq:M04 Gauss Bonnet constraint} 
\ee
Assuming validity of \eqref{eq:wrong WP form defect}, we would conclude with the help of the intersection numbers $\langle \kappa_1 \rangle=\langle \psi_i \rangle=1$ that
\be 
V_{0,4,0}(\alpha_1,\alpha_2,\alpha_3,\alpha_4)\overset{?}{=}2\pi^2 \, \Big( 1-\sum_{i=1}^4 \alpha_i^2 \Big)\ . \label{eq:four punctured sphere with defects wrong answer}
\ee
However even when obeying the constraint \eqref{eq:M04 Gauss Bonnet constraint}, it is easy to get a negative answer, e.g.\ for $\alpha_1=\alpha_2=0$, $\alpha_3=\alpha_4>\frac{1}{\sqrt{2}}$. Hence \eqref{eq:wrong WP form defect} cannot be correct.

\subsection{Sharp and blunt defects}
The reason that \eqref{eq:M04 Gauss Bonnet constraint} fails is our cavalier treatment of the boundary divisors. We have tacitly assumed that the Riemann surface under consideration is not nodal and all boundary cone points are well-separated. In fact, the gauge theory argument shows that it is true that
\be 
\frac{[\omega^\text{WP}]}{2\pi^2}=\kappa_1-\sum_i \alpha_i^2 \psi_i\quad\text{on} \quad \mathcal{M}_{g,n}\ ,
\ee
but is incorrect on $\bM_{g,n}$. 

\begin{figure}
    \centering
    \begin{tikzpicture}
        \draw[very thick, out=0, in=180] (-2,1) to (0,.5);
        \draw[very thick, out=0, in=180] (-2,-1) to (0,-.5);
        \draw[very thick, out=0, in=180] (0,.5) to (2,1);
        \draw[very thick, out=0, in=180] (0,-.5) to (2,-1);
        \draw[very thick, out=90, in=180] (-3.5,0) to (-2,1);
        \draw[very thick, out=-90, in=180] (-3.5,0) to (-2,-1);
        \draw[very thick, bend right=30] (-2.6,0.05) to (-1.4,0.05);
        \draw[very thick, bend left=30] (-2.5,0) to (-1.5,0);
        \draw[very thick, bend right=30] (1.2,0.05) to (2.8,0.05);
        \draw[very thick, bend left=30] (1.3,0) to (2.7,0);
        \draw[very thick, out=90, in=0] (3.2,0) to (2,1);
        \draw[very thick, out=-90, in=0] (3.2,0) to (2,-1);
        \draw[very thick, dashed] (0,0) ellipse (.2 and .5);
        \draw[very thick, dashed] (-3,0) ellipse (.5 and .2);
        \draw[very thick, dashed] (3,0) ellipse (.2 and .1);
        \draw[very thick, bend right=40] (-2.6,.3) to (-2.2,.8) to (-1.8,.3);
        \draw[very thick, dashed, rotate around={-30:(-1.3,.4)}] (-1.3,.4) ellipse (.2 and .43);
        \node at (-4.3,0) {$(\ell_1,\theta_1)$};
        \node at (-1.3,1.2) {$(\ell_2,\theta_2)$};
        \node at (0,.9) {$(\ell_3,\theta_3)$};
        \node at (4,0) {$(\ell_4,\theta_4)$};
    \end{tikzpicture}
    \caption{Pair of pants decomposition of a genus 2 hyperbolic surface with a cusp.}
    \label{fig:pair of pants}
\end{figure}
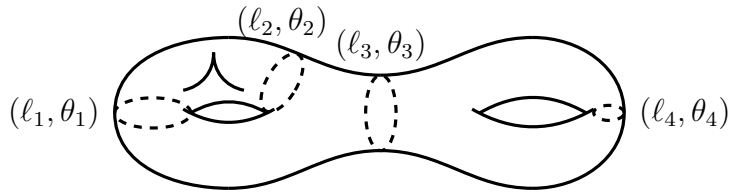
Let us first recall the situation when all defects are geodesic boundaries (or sufficiently sharp defects).
The relevant moduli space possesses a natural compactification that is constructed as follows. Every such hyperbolic surface admits a decomposition in $(3g-3+n)$ pair of pants, see Figure~\ref{fig:pair of pants} for an example. The lengths and twists $(b_i,\varrho_i)$ of this decomposition yield a natural coordinate system on (a patch of) the corresponding moduli space. Boundaries divisors of the moduli space correspond to surfaces with pinched geodesics $b_i \to 0$. This compactification precisely realizes the Deligne-Mumford compactification of moduli space. 
Wolpert's magic formula gives an explicit formula for the Weil-Petersson form in terms of Fenchel-Nielsen coordinates,
\be 
\omega_\text{WP}=\sum_i \mathrm{d}b_i \wedge \mathrm{d} \varrho_i . \label{eq:Weil-Petersson volume Fenchen-Nielsen coordinates}
\ee
This formula makes it manifest that the Weil-Petersson form extends to the compactification of moduli space, which as we already mentioned is isomorphic to the standard Deligne-Mumford compactification in the complex framework. Moreover it behaves regularly at the boundary divisors and hence the formula \eqref{eq:Weil Petersson form geodesic boundaries} does not feature any boundary classes.

The situation is very different when blunt conical defects are present, since in general there is no pair of pants decomposition. For two conical defects with deficits $\alpha_1$ and $\alpha_2$, there is only a geodesic that surrounds them when $\alpha_1+\alpha_2<1$, i.e.\ when the pair of defects is sharp. Thus when all pairs of defects are sharp, \eqref{eq:wrong WP form defect} still holds. When a pair of blunt defects is present, the structure of the compactification is different, as we already mentioned in the Introduction. The two defects will merge to a sharper conical defect with 
\be 
\alpha=1-(1-\alpha_1)-(1-\alpha_2)\ .
\ee

It is useful to generalize this scenario as follows.
Recall first that there are separating degenerating divisors $\mathscr{D}_{h,I}$ in $\bM_{g,n}$ whose Poincar\'e dual cohomology classes we denote by $\delta_{h,I}$.
Consider now the boundary divisor $\mathscr{D}_{0,I}$ of $\bM_{g,n}$ where $I \subset \{1,\dots,n\}$. A point in $\bM_{g,n}$ close to $\mathscr{D}_{0,I}$ describes a surface in which the defects defects $\alpha_i$ with $i \in I$ come close to each other. If $\alpha=1-\sum_{i \in I}(1-\alpha_i)<0$, then there is a geodesic surrounding all of the conical defects in $I$. Consequently the surface will split into two components, one spherical component containing all the conical defects in $I$ and one containing all the handles and other punctures of the surface. The two are connected at a cusp.
If on the other hand $\alpha=1-\sum_{i \in I}(1-\alpha_i)>0$, then all the conical defects can just merge into one sharper conical defect characterized by $\alpha$. As we already mentioned, this means that the Deligne-Mumford compactification is actually not the natural compactification in the present case, since it treats all conical defects as sharp. Instead, the compactification of the moduli space depends on $\boldsymbol{\alpha}=(\alpha_1,\dots,\alpha_n)$ and we denote it by $\bM_{g;\boldsymbol{\alpha}}$.

However, we can of course still give a formula for the Weil-Petersson form on all of $\bM_{g,n}$, since the Deligne-Mumford compactification is more general than the partial compactification that is adapted to the problem.

In formulas, these basic observations translate to the following. The divisor $\mathscr{D}_{0,I}$ in $\bM_{g,n}$ is isomorphic to $\bM_{0,|I|+1} \times \bM_{g,n-|I|+1}$, with the two factors describing the left and right part of the nodal surface. Let 
\be 
\xi_I: \mathscr{D}_{0,I}=\bM_{0,|I|+1} \times \bM_{g,n-|I|+1} \longrightarrow \bM_{g,n}
\ee
be the natural inclusion. Then we can consider the pullback $(\xi_I)^*([\omega_\text{WP}])$ of the Weil-Petersson class to this boundary divisor. In less fancy terms, this just means that we are restricting the symplectic form to this boundary divisor. In cohomology, this gives
\be 
(\xi_I)^*([\omega_\text{WP}])\in \H^2(\bM_{0,|I|+1}) \oplus  \H^2(\bM_{g,n-|I|+1},\RR)\ .
\ee
We can then consider the projection on the first and second factor, which we denote by $\pi^\text{L}$ and $\pi^\text{R}$. The structure of the compactification means then
\begin{subequations} \label{eq:pullback Weil Petersson form separating divisor}%
\begin{align}
    \pi^\text{L}((\xi_I)^*([\omega_\text{WP}]))&=\begin{cases}
    [\omega_\text{WP}]\ , \qquad & 1-\sum_{i \in I} (1-\alpha_i)<0\ , \\
    0\ , \qquad & 1-\sum_{i \in I} (1-\alpha_i)\ge 0\ ,
    \end{cases} \\
    \pi^\text{R}((\xi_I)^*([\omega_\text{WP}]))&=[\omega_\text{WP}]\ , 
\end{align}
\end{subequations}
where $\omega_\text{WP}$ always denoted the Weil-Petersson form on the respective space. Together with the explicit form of the Weil-Petersson form on the interior of moduli space \eqref{eq:wrong WP form defect}, we now show that this property fixes the cohomology class of the Weil-Petersson form on $\bM_{g,n}$ uniquely.
\subsection{The cohomology class of the Weil-Petersson form}\label{sec:WPform}
Using the two properties \eqref{eq:wrong WP form defect} and \eqref{eq:pullback Weil Petersson form separating divisor} we shall now demonstrate that the cohomology class of the Weil-Petersson form is completely fixed and takes the form \eqref{eq:Weil-Petersson form with boundary classes} when pulled back from $\bM_{g; \boldsymbol{\alpha}; \boldsymbol{b}}$ to $\bM_{g,|\boldsymbol{\alpha}|+|\boldsymbol{b}|}$.  

\paragraph{Uniqueness.} 
The presence of the geodesic boundaries will never influence the following proof and we can set $m=n$. Throughout we keep $\alpha_1,\dots,\alpha_n$ fixed.
The discussion so far immediately yields
\begin{align}
    \frac{[\omega_\text{WP}]}{2\pi^2}=\kappa_1-\sum_{i=1}^m \alpha_i^2 \psi_i+\sum_{\begin{subarray}{c} I \subset \{1,\dots,m\} \\ 1-\sum_{i \in I} (1-\alpha_i) \ge 0 \end{subarray}} \beta_I \, \delta_{0,I}\ , \label{eq:ansatz Weil-Petersson form}
\end{align}
since \eqref{eq:wrong WP form defect} is only modified near the boundary divisors $\mathscr{D}_{0,I}$ with $1-\sum_{i \in I} (1-\alpha_i) \ge 0$.
It remains to determine the constants $\beta_I$. For this we use the pullback property \eqref{eq:pullback Weil Petersson form separating divisor}.  With the help of the relations in Appendix~\ref{app:relation cohomology}, we can compute the pullbacks of $[\omega_\text{WP}]$ to the divisor
\be 
\mathscr{D}_{0,I} \cong \bM_{0,|I|+1}\times \bM_{g,n-|I|+1}\ .
\ee
Let's consider in particular the pullback to the projection to the second factor and denote the new marked point on $\bM_{g,n-|I|+1}$ by $\bullet$. Then $\psi_\bullet$ can only come from the pullback of the class $\delta_{0,I}$ itself, see eq.~\eqref{eq:self intersection boundary class}. Using the ansatz \eqref{eq:ansatz Weil-Petersson form}, we thus have
\be 
\pi^\text{R}\left(\xi_I^*\left(\frac{[\omega_\text{WP}]}{2\pi^2}\right)\right)=-\beta_I \psi_\bullet+ \dots\ ,
\ee
where the dots contain other linearly independent classes from $\psi_\bullet$. From our previous geometric discussion, we need this to equal the corresponding Weil-Petersson form on $\bM_{g,n-|I|+1}$, see eq.~\eqref{eq:pullback Weil Petersson form separating divisor}. There are now two cases.
\begin{enumerate}
    \item $1-\sum_{i \in I}(1-\alpha_i)<0$. In this case, the genus $g$ part of the surface has a cusp at the node $\bullet$. Since the coefficient of the $\psi$-class should vanish for a cusp, we conclude that $\beta_I=0$. This is of course consistent with the fact that the Weil-Petersson form shouldn't be modified near these boundaries and we indeed haven't included those boundary classes in the ansatz \eqref{eq:ansatz Weil-Petersson form}.
    \item $1-\sum_{i \in I}(1-\alpha_i)\ge 0$. In this case, the conical defects can merge to a new conical defect with deficit angle characterized by $\alpha=1-\sum_{i \in I}(1-\alpha_i)$. Thus the coefficient in front of $\psi_\bullet$ has to be $-\alpha^2$, which fixes 
    \be 
        \beta_I=\alpha^2=\left(1-\sum_{i \in I}(1-\alpha_i)\right)^2\ .
    \ee
\end{enumerate}
This finishes the uniqueness part of the proof.

\paragraph{Check of properties.}
We still have to demonstrate that the formula \eqref{eq:Weil-Petersson form with boundary classes}  satisfies \eqref{eq:pullback Weil Petersson form separating divisor}. Thus let us consider again the degeneration corresponding to the divisor $\mathscr{D}_{0,I}$. Without loss of generality, we can reorder the $\alpha_i$'s and assume for notational convenience that $I=\{1,\dots,\ell\}$ with $\ell \le m$. Let us distinguish again the two cases.
\begin{enumerate}
    \item $1-\sum_{i=1}^\ell(1-\alpha_i)<0$. In this case, the boundary class $\delta_{0,I}$ itself is absent and there is in particular no self-intersection. $\kappa_1$ pulls back to the sum of the $\kappa_1$-classes on $\bM_{0,\ell+1} \times \bM_{g,n-\ell+1}$. The first $\ell$ $\psi$-classes pullback to the corresponding $\psi$-classes on $\bM_{0,\ell+1}$ (with the $\psi$-class at the cusp absent) and similarly the other $\psi$-classes pullback to the corresponding $\psi$-classes on $\bM_{g,n-\ell+1}$. Finally, the $\delta$-classes $\delta_{0,J}$ with $J \subsetneq I$ pullback to the corresponding $\delta$-classes on $\bM_{0,\ell+1}$ and the $\delta$-classes with $J \cap I=\emptyset$ pull back to the corresponding $\delta$-classes on $\bM_{g,n-\ell+1}$. $\delta$-classes with $I \subsetneq J$ do not appear because of our assumption that $1-\sum_{i \in J}(1-\alpha_i)<0$. All other $\delta$-classes have zero pullback. These properties make it obvious that the pullbacks to both components lead to the correct Weil-Petersson class as predicted from the equation \eqref{eq:Weil-Petersson form with boundary classes}.
    \item $1-\sum_{i=1}^\ell(1-\alpha_i)\ge 0$. Let us denote the new $\psi$-classes at the node for the left (i.e.\ $\bM_{0,\ell+1}$) and right (i.e.\ $\bM_{g,n-\ell+1}$) component by $\psi_\circ$ and $\psi_\bullet$. The pullbacks of $\kappa_1$ and the $\psi$-classes are unchanged from the previous case, but there are new contributions from the $\delta$-classes. We first have the contribution from the self-pullback of $\delta_{0,I}$ which gives (see eq.~\eqref{eq:self intersection boundary class})
    \be 
        \left(1-\sum_{i=1}^\ell (1-\alpha_i)\right)^2 (-\psi_\circ-\psi_\bullet)\ .   
    \ee
    There are also new contributions of pullbacks from $\delta$-classes to the right component. We have
    \be 
        \pi^\text{R}(\xi_{0,I}^*(\delta_{0,J}))=\delta_{0, \{\bullet\} \cup J \setminus I}    
    \ee
    when $I \subsetneq J$. These terms are precisely needed to make the projection to the right factor work, since they supply the $\delta$-classes needed for when the new merged conical defect collides with other conical defects, which can happen when $1-\sum_{i \in J}(1-\alpha_i)\ge 0$ with $I \subsetneq J$. Thus these two additional terms precisely make the pullback to the right component work.
    
    It remains to show that the pullback to the left component is in fact zero. From our discussion so far, it equals
    \begin{align}
        \pi^\text{L}\left(\xi_I^*\left(\frac{[\omega_\text{WP}]}{2\pi^2}\right)\right)&=\kappa_1-\sum_{i=1}^\ell \alpha_i^2 \psi_i-\left(1-\sum_{i=1}^\ell(1-\alpha_i)\right)^2 \psi_\circ\nonumber\\
        &\qquad+\sum_{J \subsetneq I} \left(1-\sum_{i\in J}(1-\alpha_i)\right)^2 \delta_{0,J}\ .
    \end{align}
    To show that this is in fact zero, we can use the additional relations for these classes in $\H^2(\bM_{0,\ell+1},\RR)$, which we spelled out in Appendix~\ref{app:relation cohomology}. 
    Let's proceed coefficient by coefficient. To simplify the notation, we always assume that sums over subsets $J$ run over strict subsets $J \subsetneq \{1,\dots,\ell\}$.
    
    Using eq.~\eqref{eq:psi class identity M0n}, the coefficient of $\alpha_i^2$ equals
    \be 
        -\psi_i+\psi_\circ+\sum_{J ,\ i \in J} \delta_{0,J}=-\sum_{J,\ i \in J,\, j \not\in J} \delta_{0,J} -\sum_{J, \ i,\, j \not \in J} \delta_{0,J \cup \{ \circ\}} +\sum_{J,\ i \in J} \delta_{0,J}\ ,
    \ee
    where $j$ is an index not equal to $i$ and $\circ$. We can replace the second sum with the sum over its complement in $\{1,\dots,\ell,\circ\}$, which combines with the first term to cancel the third term. 
    Next, the coefficient of $\alpha_i \alpha_j$ equals
    \begin{align}
        -2 \psi_\circ+2\sum_{J,\ i,\, j \in J} \delta_{0,J}\ . \label{eq:coefficient alphai alphaj}
    \end{align}
    We can again use eq.~\eqref{eq:psi class identity M0n} to express in $\psi_\circ$ in terms of boundary classes. After replacing the sum with the corresponding sum over the complement, it precisely cancels the second term in this expression.
    The coefficient of $\alpha_i$ equals
    \begin{align}
        -2(\ell-1)\psi_\circ+2\sum_{J,\, i \in J} (|J|-1)\,  \delta_{0,J}\ . \label{eq:coefficient alphai}
    \end{align}
    This is the same as summing \eqref{eq:coefficient alphai alphaj} over $j\in I=\{1,\dots,\ell\}$ and hence also vanishes.
    Finally, the coefficient of the constant term equals
    \begin{align}
        \kappa_1-(\ell-1)^2 \psi_\circ+\sum_{J} (|J|-1)^2\,  \delta_{0,J}\ . \label{eq:coefficient 1}
    \end{align}
    We use \eqref{eq:kappa class identity M0n} with $j=\circ$ to express $\kappa_1$ in terms of boundary classes,
    \be 
       \kappa_1=\sum_{J,\, i \not\in J}     (|J|-1)\, \delta_{0,J}\ .
    \ee
    Since this identity is valid for any choice of $i \in \{1,\dots,\ell\}$, we can sum over the choice of $i$, which gives
    \be 
        \ell \kappa_1=\sum_{J} (|J|-1)(\ell-|J|)\,  \delta_{0,J}\ .    
    \ee
    Similarly, we can use that \eqref{eq:coefficient alphai} vanishes and sum over the choice of $i \in \{1,\dots,\ell\}$, which gives
    \be
        \ell(\ell-1)\psi_\circ=\sum_{J} |J|(|J|-1)\,   \delta_{0,J}\ .
    \ee
    We thus get for the coefficient of the constant term \eqref{eq:coefficient 1}
    \begin{align}
       \frac{1}{\ell} \sum_{J} \left((|J|-1)(\ell-|J|) -(|J|-1) |J|(\ell-1)+\ell(|J|-1)^2\right)\,\delta_{0,J}=0\ .
    \end{align}
\end{enumerate}
This finishes the demonstration that $[\omega^\text{WP}]$ as given by \eqref{eq:Weil-Petersson form with boundary classes} satisfies the desired pullback properties \eqref{eq:pullback Weil Petersson form separating divisor}.
\subsection{The dilaton operator}
We now establish some further properties of our formula for the cohomology class of the Weil-Petersson form. The two equations that we derive are the generalizations of the string and the dilaton equation and are stated in \eqref{eq:string equation} and \eqref{eq:dilaton equation}. We assume again that punctures $1,\dots,m$ are conical defects and punctures $m+1,\dots,n$ are geodesic boundaries.

Let us start by computing the pullback of $[\omega_\text{WP}]$ under the forgetful morphism
\be 
    \pi:\bM_{g,n+1} \longrightarrow \bM_{g,n}\ ,
\ee
that forgets the $(m+1)$-st puncture.
With the help of the equations \eqref{eq:pullback classes forgetful morphism}, we get
\begin{align}
    \pi^*\left(\frac{[\omega_\text{WP}]}{2\pi^2}\right)&=\kappa_1-\psi_{m+1}-\sum_{i=1}^m \alpha_i^2\psi_i+\frac{1}{4\pi^2}\sum_{i=1}^{n-m} b_i^2 \psi_{m+i+1}+\sum_{i=1}^m \alpha_i^2 \delta_{0,\{i,m+1\}}\nonumber\\
    &\qquad+\sum_{\begin{subarray}{c} I \subset \{1,\dots,m\} \\ 1-\sum_{i \in I} (1-\alpha_i) \ge 0 \end{subarray}} \left(1-\sum_{i \in I} (1-\alpha_i) \right)^2 \left(\delta_{0,I}+\delta_{0,I\cup \{m+1\}}\right)\nonumber\\
    &\qquad-\frac{1}{4\pi^2} \sum_{i=1}^{n-m} b_i^2 \delta_{0,\{i+m+1,m+1\}}\ .
\end{align}
Apart from the term in the last line, we recognize this to be the Weil-Petersson form with one additional conical defect with zero defect angle (i.e.\ $\alpha_{m+1}=1$) inserted. The term in the last line is absent on a surface without geodesic boundaries. This is geometrically of course completely obvious; when adding a further marked point to the surface without geodesic boundaries nothing special happens there and it corresponds to a further `defect' with zero deficit angle, i.e.\ no defect at all. The correction with geodesic boundaries roughly has the interpretation of accounting for the possibility that the geodesic boundaries introduce a hole in the surface of finite size and the `defect' is not allowed to end up in the hole, i.e.\ outside the surface.

We thus have
\be 
\pi^*([\omega_\text{WP}])=[\omega_\text{WP}]-\frac{1}{2} \sum_{i=1}^{n-m} b_i^2 \delta_{0,\{i+m+1,m+1\}}\ ,
\ee
where the Weil-Petersson form on the right has a zero defect angle `defect' angle inserted. 
With the preparations, we can now derive \eqref{eq:string equation} and \eqref{eq:dilaton equation}.

\paragraph{String equation.} The left-hand side of \eqref{eq:string equation} is computed by
\be 
    V_{g,m+1,n-m}(\boldsymbol{\alpha},1;\boldsymbol{b})
    = \left\langle \exp\left(\pi^*([\omega_\text{WP}])+\frac{1}{2}\sum_{i=1}^{n-m} b_i^2 \delta_{0,\{i+m+1,m+1\}}\right) \right \rangle \ ,
\ee
where we adopted the standard convention that the angle bracket denotes the integral over the respective moduli space. 
We can now expand the second term in the exponential as a power series. Since $\delta_{0,\{i+m+1,m+1\}}\delta_{0,\{j+m+1,m+1\}}=0$ for $i \ne j$ because the corresponding divisors don't intersect, we do not have to keep any cross terms in the expansion. We also don't have to keep the zeroth order term because
\be 
\left\langle (\pi^*([\omega_\text{WP}]))^{3g-2+n} \right\rangle=
\left\langle \pi^*\left([\omega_\text{WP}]^{3g-2+n}\right) \right\rangle=0\ ,
\ee
since $[\omega_\text{WP}]^{3g-2+n}$ vanishes for dimensional reasons. We thus get
\begin{align}
    V_{g,m+1,n-m}(\boldsymbol{\alpha},1;\boldsymbol{b})=\sum_{i=1}^{n-m}\sum_{k\ge 1} \frac{1}{k!} \left(\frac{b_i^2}{2}\right)^k \left \langle \delta_{0,\{i+m+1,m+1\}}^k \exp \left(\pi^*([\omega_\text{WP}])\right)\right \rangle \ .
\end{align}
We can evaluate the relevant intersection numbers with the boundary classes by pulling them back to the boundary divisor. Let $\xi_{0,\{i+m+1,m+1\}}$ be again the inclusion of the boundary divisor $\mathscr{D}_{0,\{i+m+1,m+1\}} \hookrightarrow \bM_{g,n+1}$. For $k=1$, we have
\begin{align}
    \left \langle \delta_{0,\{i+m+1,m+1\}} \exp \left(\pi^*([\omega_\text{WP}])\right)\right \rangle&=\left\langle\xi_{0,\{i+m+1,m+1\}}^* \pi^*\exp \left([\omega_\text{WP}]\right)\right \rangle\\
    &=\left\langle \exp\left([\omega_\text{WP}]\right) \right \rangle \ ,
\end{align}
where used that $\pi \circ \xi_{0,\{i+m+1,m+1\}}$ is the identity map on $\bM_{g,n}$ and hence the composition $\xi_{0,\{i+m+1,m+1\}}^* \pi^*$ is also the identity in cohomology. For higher $k$, we have
\begin{align}
    \left \langle \delta_{0,\{i+m+1,m+1\}}^k \exp \left(\pi^*([\omega_\text{WP}])\right)\right \rangle&=\left\langle\xi_{0,\{i+m+1,m+1\}}^* \left(\delta_{0,\{i+m+1,m+1\}}^{k-1}\pi^*\exp \left([\omega_\text{WP}]\right)\right)\right \rangle \\
    &=\left\langle (-\psi_{i+m+1})^{k-1} \exp \left([\omega_\text{WP}]\right)\right \rangle \ ,
\end{align}
where we used \eqref{eq:self intersection boundary class} for the relevant pullback. We thus have
\begin{align}
    V_{g,m+1,n-m}(\boldsymbol{\alpha},1;\boldsymbol{b})&=\sum_{i=1}^{n-m}\sum_{k\ge 1} \frac{1}{k!} \left(\frac{b_i^2}{2}\right)^k \left\langle (-\psi_{i+m+1})^{k-1} \exp \left([\omega_\text{WP}]\right)\right \rangle \\
    &=\sum_{i=1}^{n-m} \left\langle \frac{1-\e^{-\frac{1}{2}b_i^2 \psi_{i+m+1}}}{\psi_{i+m+1}} \exp \left([\omega_\text{WP}]\right)\right \rangle \ .
\end{align}
Noting that
\be 
\frac{1-\e^{-\frac{1}{2}b_i^2 \psi_{i+m+1}}}{\psi_{i+m+1}} \, \e^{\frac{1}{2}b_i^2 \psi_{i+m+1}}=\int\mathrm{d}b_i \, b_i\  \e^{\frac{1}{2}b_i^2 \psi_{i+m+1}}\ ,
\ee
the string equation now follows.

\paragraph{Dilaton equation.}
The dilaton equation is similar to demonstrate. We can compute
\begin{multline}
    \frac{\mathrm{d}V_{g,m+1,n-m}(\boldsymbol{\alpha},\alpha_{m+1};\boldsymbol{b})}{\mathrm{d}\alpha_{m+1}}\, \Big|_{\alpha_{m+1}=1}\\
    =\left\langle \exp\left(\pi^*([\omega_\text{WP}])+\frac{1}{2}\sum_{i=1}^{n-m} b_i^2\,  \delta_{0,\{i+m+1,m+1\}}\right) \frac{\mathrm{d}[\omega_\text{WP}] }{\mathrm{d}\alpha_{m+1}} \Big|_{\alpha_{m+1}=1} \right \rangle \ .
\end{multline}
We can integrate over the fiber by applying the pushforward $\pi_*$ and using the basic property $\pi_* \left(\pi^* \alpha \, \beta\right)=\alpha\, \pi_*(\beta)$ of the pushforward. This gives
\begin{multline}
    \frac{\mathrm{d}V_{g,m+1,n-m}(\boldsymbol{\alpha},\alpha_{m+1};\boldsymbol{b})}{\mathrm{d}\alpha_{m+1}}\, \Big|_{\alpha_{m+1}=1}\\
    =\left\langle \exp\left([\omega_\text{WP}]\right)\ \pi_*\left(\exp\left(\frac{1}{2}\sum_{i=1}^{n-m} b_i^2\,  \delta_{0,\{i+m+1,m+1\}}\right)\frac{\mathrm{d} [\omega_\text{WP}]}{\mathrm{d}\alpha_{m+1}}  \Big|_{\alpha_{m+1}=1}\right) \right \rangle\ .
\end{multline}
Let us compute the relevant pushforward. Notice first that 
\be 
\frac{\mathrm{d} [\omega_\text{WP}]}{\mathrm{d}\alpha_{m+1}}  \Big|_{\alpha_{m+1}=1}=-2 \psi_{m+1}+2\sum_{I \subset \{1,\dots,m+1\},\, m+1 \in I} \left(1-\sum_{i \in I} (1-\alpha_i)\right) \delta_{0,I}\ .
\ee
All the appearing classes vanish when multiplied with $\delta_{0,\{i+m+1,m+1\}}$. For the boundary classes, this is geometrically obvious because the divisors $\mathscr{D}_{0,\{i+m+1,m+1\}}$ and $\mathscr{D}_{0,I}$ with $I \subset \{1,\dots,m\}$ do not intersect. For the $\psi_{m+1}$-class, this follows because the defining line bundle $\LL_{m+1}$ is trivial when restricted to $\mathscr{D}_{0,\{i+m+1,m+1\}}$. Thus, we may omit the exponential factor and it remains to compute the pushforward of $\frac{\mathrm{d}[\omega_\text{WP}]}{\mathrm{d}\alpha_{m+1}}  \Big|_{\alpha_{m+1}=1}$. This can be done with the help of the pushforwards given in Appendix~\ref{app:relation cohomology},
\begin{align}
    \pi_*\left(\frac{\mathrm{d}[\omega_\text{WP}]}{\mathrm{d}\alpha_{m+1}}  \Big|_{\alpha_{m+1}=1}\right)=2\left(-2g+2-n+\sum_{i=1}^m \alpha_i\right)\equiv 2\,\chi(\Sigma_{g,\boldsymbol{\alpha},\boldsymbol{b}})\ .
\end{align}
This proves the dilaton equation.

\subsection{Some simple examples}
Let us list some WP-volumes with conical defects. We restrict ourselves to only defects and no geodesic boundaries. We always assume that they are ordered such that $\alpha_1 \le \alpha_2 \le \dots \le \alpha_n$.
\paragraph{$g=0$, $n=4$.} Let us denote
\be 
\theta(x)=x^2 \Theta(x)\ ,
\ee
where $\Theta(x)$ is the Heaviside theta function.
Using $\langle \kappa_1 \rangle=\langle \psi_i \rangle=1$ on $\bM_{0,4}$, we get
\begin{align}
    \frac{V_{0,4,0}(\boldsymbol{\alpha})}{2\pi^2}&=1-\sum_{i=1}^4 \alpha_i^2+\sum_{1 \le i< j\le 4} \theta(\alpha_i+\alpha_j-1) \\
    &=\begin{cases}
    1-\alpha_1^2-\alpha_2^2-\alpha_3^2-\alpha_4^2\ , \\
    -\alpha_1^2-\alpha_2^2+2(1-\alpha_3)(1-\alpha_4)\ , &\ \alpha_3+\alpha_4 > 1\ ,  \\
    -\alpha_1^2+(1-\alpha_4) (3-2 \alpha_2-2 \alpha_3-\alpha_4)\ , &\ \alpha_2+\alpha_4>1\ , \\
    -\alpha_1^2+(2-\alpha_2-\alpha_3-\alpha_4)^2\ , &\ \alpha_2+\alpha_3>1\ , \\
    2(1-\alpha_4)(2-\alpha_1-\alpha_2-\alpha_3-\alpha_4)\ , &\ \alpha_1+\alpha_4>1\ .
    \end{cases} \label{eq:V04 cases}
\end{align}
It is of course understood that the Gauss-Bonnet constraint \eqref{eq:M04 Gauss Bonnet constraint} is satisfied. 
The conditions are not mutually exclusive. 
In order to not write too many conditions, it is also understood that these conditions are to be read sequentially and one should pick the last condition in the list that applies.

Geometrically, these conditions mean the following. In the first case, all pairs of defects are sharp. In the second case, there is exactly one pair of blunt defects, namely 3 and 4. In the third case, there is a `chain' of three defects that can merge, i.e.\ the first can merge with the second and the second can merge with the third, but not the first with the third. Since the $\alpha_i$'s are assumed to be ordered, defects 2, 3 and 4 form this chain. In the fourth case, there is a triangle of defects that can merge and in the fifth case, the pairs $(1,4)$, $(2,4)$ and $(3,4)$ are blunt. We could graphically denote the second through fifth case by the graphs
\begin{align}
    \begin{tikzpicture}[baseline={([yshift=-.5ex]current bounding box.center)}]
    \draw[very thick] (0,0) -- (1,0);
    \fill (0,0) circle (.07) node[below] {3};
    \fill (1,0) circle (.07) node[below] {4};
    \end{tikzpicture}\ , \quad  
    \begin{tikzpicture}[baseline={([yshift=-.5ex]current bounding box.center)}]
    \draw[very thick] (0,0) -- (2,0);
    \fill (0,0) circle (.07) node[below] {2};
    \fill (1,0) circle (.07) node[below] {4};
    \fill (2,0) circle (.07) node[below] {3};
    \end{tikzpicture}\ ,  \quad 
    \begin{tikzpicture}[baseline={([yshift=-.5ex]current bounding box.center)}]
    \draw[very thick] (0,0) -- (1,0) -- (.5,.87) -- (0,0);
    \fill (0,0) circle (.07) node[below] {2};
    \fill (1,0) circle (.07) node[below] {3};
    \fill (.5,.87) circle (.07) node[above] {4};
    \end{tikzpicture}\ , \quad 
        \begin{tikzpicture}[baseline={([yshift=-.5ex]current bounding box.center)}]
    \draw[very thick] (0,0) -- (1,0);
    \draw[very thick] (0,0) -- (-.5,.87);
    \draw[very thick] (0,0) -- (-.5,-.87);
    \fill (0,0) circle (.07) node[left] {4};
    \fill (1,0) circle (.07) node[below] {1};
    \fill (-.5,.87) circle (.07) node[left] {2};
    \fill (-.5,-.87) circle (.07) node[left] {3};
    \end{tikzpicture}\ ,
\end{align}
where a line connecting two nodes means that this pair of indices is blunt.

Contrary to the naive answer \eqref{eq:four punctured sphere with defects wrong answer}, it is easy to check that this answer is always positive. Note also that the last case in \eqref{eq:V04 cases} is consistent and completely specified by the string and dilaton equations \eqref{eq:string and dilaton equations}.
\paragraph{$g=0$, $n=5$.} In the presence of five defects, there are many cases. We can denote them again graphically. There are 16 basic cases, which correspond to the graphs
\begin{align}
   & \emptyset \ , \quad 
     \begin{tikzpicture}[baseline={([yshift=-.5ex]current bounding box.center)}]
    \draw[very thick] (0,0) -- (1,0);
    \fill (0,0) circle (.07) node[below] {4};
    \fill (1,0) circle (.07) node[below] {5};
    \end{tikzpicture}\ , \quad  
    \begin{tikzpicture}[baseline={([yshift=-.5ex]current bounding box.center)}]
    \draw[very thick] (0,0) -- (2,0);
    \fill (0,0) circle (.07) node[below] {3};
    \fill (1,0) circle (.07) node[below] {5};
    \fill (2,0) circle (.07) node[below] {4};
    \end{tikzpicture}\ ,  \quad 
        \begin{tikzpicture}[baseline={([yshift=-.5ex]current bounding box.center)}]
    \draw[very thick] (0,0) -- (1,0);
    \draw[very thick] (0,0) -- (-.5,.87);
    \draw[very thick] (0,0) -- (-.5,-.87);
    \fill (0,0) circle (.07) node[left] {5};
    \fill (1,0) circle (.07) node[below] {2};
    \fill (-.5,.87) circle (.07) node[left] {3};
    \fill (-.5,-.87) circle (.07) node[left] {4};
    \end{tikzpicture}\ , \quad 
    \begin{tikzpicture}[baseline={([yshift=-.5ex]current bounding box.center)}]
    \draw[very thick] (0,0) -- (1,0) -- (.5,.87) -- (0,0);
    \fill (0,0) circle (.07) node[below] {3};
    \fill (1,0) circle (.07) node[below] {4};
    \fill (.5,.87) circle (.07) node[above] {5};
    \end{tikzpicture}\ , \quad 
    \begin{tikzpicture}[baseline={([yshift=-.5ex]current bounding box.center)}]
    \draw[very thick] (0,0) -- (1,0);
    \draw[very thick] (0,0) -- (-1,0);
    \draw[very thick] (0,0) -- (0,1);
    \draw[very thick] (0,0) -- (0,-1);
    \fill (0,0) circle (.07);
    \node at (.25,.25)  {5};
    \fill (1,0) circle (.07) node[below] {1};
    \fill (0,1) circle (.07) node[left] {2};
    \fill (-1,0) circle (.07) node[below] {3};
    \fill (0,-1) circle (.07) node[left] {4};
    \end{tikzpicture}\ , \nonumber\\
    &\begin{tikzpicture}[baseline={([yshift=-.5ex]current bounding box.center)}]
    \draw[very thick] (0,0) -- (.87,-.5) -- (.87,.5) -- (0,0) ;
    \draw[very thick] (0,0) -- (-1,0);
    \fill (0,0) circle (.07) node[below] {5};
    \fill (.87,.5) circle (.07) node[right] {3};
    \fill (.87,-.5) circle (.07) node[right] {4};
    \fill (-1,0) circle (.07) node[below] {2};
    \end{tikzpicture}\ , \quad 
    \begin{tikzpicture}[baseline={([yshift=-.5ex]current bounding box.center)}]
    \draw[very thick] (0,0) -- (.87,-.5) -- (.87,.5) -- (0,0) ;
    \draw[very thick] (0,0) -- (-.87,.5);
    \draw[very thick] (0,0) -- (-.87,-.5);
    \fill (0,0) circle (.07) node[below] {5};
    \fill (.87,.5) circle (.07) node[right] {3};
    \fill (.87,-.5) circle (.07) node[right] {4};
    \fill (-.87,-.5) circle (.07) node[left] {2};
    \fill (-.87,.5) circle (.07) node[left] {3};
    \end{tikzpicture}\ , \quad 
    \begin{tikzpicture}[baseline={([yshift=-.5ex]current bounding box.center)}]
    \draw[very thick] (0,.5) -- (.87,0) -- (0,-.5) -- (-.87,0) -- (0,.5);
    \draw[very thick] (0,.5) -- (0,-.5);
    \fill (0,.5) circle (.07) node[above] {5};
    \fill (.87,0) circle (.07) node[right] {2};
    \fill (-.87,0) circle (.07) node[left] {3};
    \fill (0,-.5) circle (.07) node[below] {4};
    \end{tikzpicture}\ , \quad 
    \begin{tikzpicture}[baseline={([yshift=-.5ex]current bounding box.center)}]
    \draw[very thick] (.5,0) -- (0,.87) -- (-.5,0) -- (0,-.87) -- (.5,0);
    \draw[very thick] (.5,0) -- (-1.5,0);
    \fill (.5,0) circle (.07) node[right] {4};
    \fill (0,.87) circle (.07) node[right] {2};
    \fill (0,-.87) circle (.07) node[right] {3};
    \fill (-1.5,0) circle (.07) node[below] {1};
    \fill (-.5,0) circle (.07);
    \node at (-.7,.25) {5};
    \end{tikzpicture}\ , \nonumber\\
    &\begin{tikzpicture}[baseline={([yshift=-.5ex]current bounding box.center)}]
    \draw[very thick] (0,0) -- (1,0) -- (1,1) -- (0,1) -- (0,0);
    \draw[very thick] (0,0) -- (1,1);
    \draw[very thick] (0,1) -- (1,0);
    \fill (0,0) circle (.07) node[below] {5};
    \fill (0,1) circle (.07) node[above] {4};
    \fill (1,0) circle (.07) node[below] {3};
    \fill (1,1) circle (.07) node[above] {2};
    \end{tikzpicture}\ , \quad 
    \begin{tikzpicture}[baseline={([yshift=-.5ex]current bounding box.center)}]
    \draw[very thick] (0,-.5) -- (0,.5) -- (.87,0) --(0,-.5);
    \draw[very thick] (0,-.5) -- (-1,.5) -- (0,.5);
    \draw[very thick] (0,.5) -- (-1,-.5) -- (0,-.5);
    \fill (0,.5) circle (.07) node[above] {5};
    \fill (0,-.5) circle (.07) node[below] {4};
    \fill (.87,0) circle (.07) node[right] {1};
    \fill (-1,.5) circle (.07) node[above] {3};
    \fill (-1,-.5) circle (.07) node[below] {2};
    \end{tikzpicture}\ , \quad 
    \begin{tikzpicture}[baseline={([yshift=-.5ex]current bounding box.center)}]
    \draw[very thick] (0,0) -- (.71,.71) -- (1.42,0) -- (.71,-.71) -- (0,0) -- (-1,0);
    \draw[very thick] (.71,.71) -- (.71,-.71);
    \draw[very thick] (0,0) -- (1.42,0);
    \fill (-1,0) circle (.07) node[below] {1};
    \fill (.71,.71) circle (.07) node[above] {3};
    \fill (1.42,0) circle (.07) node[right] {2};
    \fill (.71,-.71) circle (.07) node[below] {4};
    \fill (0,0) circle (.07);
    \node at (-.2,.25) {5};
    \end{tikzpicture}\ , \quad 
    \begin{tikzpicture}[baseline={([yshift=-.5ex]current bounding box.center)}]
    \draw[very thick] (0,-.5) -- (0,.5) -- (.87,0) --(0,-.5);
    \draw[very thick] (0,-.5) -- (-1,.5) -- (0,.5);
    \draw[very thick] (0,.5) -- (-1,-.5) -- (0,-.5);
    \draw[very thick] (-1,.5) -- (-1,-.5);
    \fill (0,.5) circle (.07) node[above] {5};
    \fill (0,-.5) circle (.07) node[below] {4};
    \fill (.87,0) circle (.07) node[right] {1};
    \fill (-1,.5) circle (.07) node[above] {3};
    \fill (-1,-.5) circle (.07) node[below] {2};
    \end{tikzpicture}\ , \nonumber \\
    &\begin{tikzpicture}[baseline={([yshift=-.5ex]current bounding box.center)}]
    \draw[very thick] (0,0) -- (1,1) -- (2,0) -- (1,-1) -- (0,0);
    \draw[very thick] (1,1) -- (1,-1);
    \draw[very thick] (0,0) -- (2,0);
    \draw[very thick, bend left=30] (1,1) to (1,-1);
    \fill (0,0) circle (.07) node[left] {1};
    \fill (1,1) circle (.07) node[above] {3};
    \fill (2,0) circle (.07) node[right] {2};
    \fill (1,-1) circle (.07) node[below] {5};
    \fill (1,0) circle (.07);
    \node at (.8,.25) {4};
    \end{tikzpicture}\ , \quad 
    \begin{tikzpicture}[baseline={([yshift=-.5ex]current bounding box.center)}]
    \draw[very thick] (1,0) -- (-.81,.59) -- (.31,-.95) -- (.31,.95) -- (-.81,-.59) -- (1,0);
    \draw[very thick] (1,0) -- (.31,.95) -- (-.81,.59) -- (-.81,-.59) -- (.31,-.95) -- (1,0);
    \fill (1,0) circle (.07) node[right] {1};
    \fill (.31,.95) circle (.07) node[above] {2};
    \fill (-.81,.59) circle (.07) node[left] {3};
    \fill (-.81,-.59) circle (.07) node[left] {4};
    \fill (.31,-.95) circle (.07) node[below] {5};
    \end{tikzpicture}\ .     \label{eq:M05 defect possibilities}
\end{align}
Additionally, most of these cases have further subcases. For five defects it is sometimes also allowed that three of them merge simultaneously and hence a boundary class of the form $\delta_{0,\{i,j,k\}}$ is present. We can denote this by shading the corresponding face of the graph. For example, the fifth case above has two subcases, which are
\begin{align}
    \begin{tikzpicture}[baseline={([yshift=-.5ex]current bounding box.center)}]
    \draw[very thick] (0,0) -- (1,0) -- (.5,.87) -- (0,0);
    \fill (0,0) circle (.07) node[below] {3};
    \fill (1,0) circle (.07) node[below] {4};
    \fill (.5,.87) circle (.07) node[above] {5};
    \end{tikzpicture}\quad\text{and} \quad 
    \begin{tikzpicture}[baseline={([yshift=-.5ex]current bounding box.center)}]
    \fill[black!30!white] (0,0) -- (1,0) -- (.5,.87) -- (0,0);
    \draw[very thick] (0,0) -- (1,0) -- (.5,.87) -- (0,0);
    \fill (0,0) circle (.07) node[below] {3};
    \fill (1,0) circle (.07) node[below] {4};
    \fill (.5,.87) circle (.07) node[above] {5};
    \end{tikzpicture}\ .
\end{align}
In total there are 66 cases that arise from shading the above diagrams \eqref{eq:M05 defect possibilities} appropriately.
For a subset $I \subset \{1,2,3,4,5\}$, let us write
\be 
\theta_I(\boldsymbol{\alpha})=\theta\left(1-\sum_{i\in I} (1-\alpha_i)\right)\ .
\ee
Then the class of the Weil-Petersson form can be written as
\begin{align}
    \frac{[\omega_\text{WP}]}{2\pi^2}=\kappa_1-\sum_i \alpha_i^2 \psi_i+\sum_{I \subset \{1,2,3,4,5\},\, |I|=2} \big(\theta_I(\boldsymbol{\alpha})+\theta_{I^c}(\boldsymbol{\alpha})\big) \, \delta_{0,I}\ .
\end{align}
With the help of the identities in Appendix~\ref{app:intersection numbers}, we then compute
\begin{align}
    \frac{V_{0,5,0}(\boldsymbol{\alpha})}{(2\pi^2)^2}&=\frac{ \langle \kappa_1^2 \rangle}{2}+\frac{1}{2}\sum_i \alpha_i^4 \langle \psi_i^2 \rangle+\sum_{i<j} \alpha_i^2 \alpha_j^2 \langle \psi_i \psi_j \rangle-\sum_i \alpha_i^2 \langle \kappa_1 \psi_i \rangle \nonumber\\
    &\qquad+\sum_{I} \big(\theta_I(\boldsymbol{\alpha})+\theta_{I^c}(\boldsymbol{\alpha})\big) \langle \delta_{0,I} \kappa_1 \rangle-\sum_i \sum_I \alpha_i^2 \big(\theta_I(\boldsymbol{\alpha})+\theta_{I^c}(\boldsymbol{\alpha})\big) \langle \psi_i \delta_{0,I} \rangle\nonumber\\
    &\qquad+\frac{1}{2} \sum_{I,J} \big(\theta_I(\boldsymbol{\alpha})+\theta_{I^c}(\boldsymbol{\alpha})\big)\big(\theta_J(\boldsymbol{\alpha})+\theta_{J^c}(\boldsymbol{\alpha})\big) \langle \delta_{0,I} \delta_{0,J} \rangle \\
    &=\frac{5}{2}+\frac{1}{2} \sum_i \alpha_i^4+2 \sum_{i<j} \alpha_i^2 \alpha_j^2-3 \sum_i \alpha_i^2+\sum_I \big(\theta_I(\boldsymbol{\alpha})+\theta_{I^c}(\boldsymbol{\alpha})\big)\nonumber\\
    &\qquad-\sum_I \sum_{j \not \in I} \alpha_j^2 \big(\theta_I(\boldsymbol{\alpha})+\theta_{I^c}(\boldsymbol{\alpha})\big)-\frac{1}{2} \sum_I \big(\theta_I(\boldsymbol{\alpha})+\theta_{I^c}(\boldsymbol{\alpha})\big)^2\nonumber\\
    &\qquad+\frac{1}{2}\sum_{I,J ,\,  I \cap J=\emptyset} \big(\theta_I(\boldsymbol{\alpha})+\theta_{I^c}(\boldsymbol{\alpha})\big)\big(\theta_J(\boldsymbol{\alpha})+\theta_{J^c}(\boldsymbol{\alpha})\big)\ .
\end{align}
Here sums over subset $I$ are always understood to be over subsets $I \subset\{1,2,3,4,5\}$ with $|I|=2$. One can also check that this formula is consistent with the string and dilaton equations and always gives a positive result.

We can simplify this result as follows. Let us collect all the $\theta_{I^c}$ together and sum over the complement instead. Notice that $\theta_I(\boldsymbol{\alpha}) \theta_{I^c}(\boldsymbol{\alpha})=0$ because of the Gauss-Bonnet constraint. Notice also that
\be 
\sum_{I,J,|I|=2,|J|=2,I \cap J=\emptyset}\theta_{I^c}(\boldsymbol{\alpha})\theta_J(\boldsymbol{\alpha})=\sum_{I,|I|=3} \sum_{j,k \in I, j<k} (\alpha_j+\alpha_k-1)^2\theta_I(\boldsymbol{\alpha}) 
\ee
These terms are
\begin{multline}
    \sum_{I,|I|=3} \left[\Big(1-\sum_{i \in I} \alpha_i^2\Big)\theta_I(\boldsymbol{\alpha})-\frac{1}{2} \theta_I(\boldsymbol{\alpha})^2+\sum_{j,k \in I, j<k} (\alpha_j+\alpha_k-1)^2 \theta_I(\boldsymbol{\alpha})\right]\\
    =\frac{1}{2} \sum_{I,|I|=3} \theta_I(\boldsymbol{\alpha})^2\ .
\end{multline}
Thus we obtain
\begin{align}
    \frac{V_{0,5,0}(\boldsymbol{\alpha})}{(2\pi^2)^2}
    &=\frac{5}{2}+\frac{1}{2} \sum_i \alpha_i^4+2 \sum_{i<j} \alpha_i^2 \alpha_j^2-3 \sum_i \alpha_i^2+\sum_{I,|I|=2} \theta_I(\boldsymbol{\alpha})\nonumber\\
    &\qquad-\sum_{I,|I|=2} \sum_{j \not \in I} \alpha_j^2 \theta_I(\boldsymbol{\alpha})-\frac{1}{2} \sum_{I,|I|=2} \theta_I(\boldsymbol{\alpha})^2\nonumber\\
    &\qquad+\frac{1}{2}\sum_{I,J ,|I|=|J|=2,\,  I \cap J=\emptyset} \theta_I(\boldsymbol{\alpha})\theta_J(\boldsymbol{\alpha})+\frac{1}{2}\sum_{I,|I|=3} \theta_I(\boldsymbol{\alpha})^2\ .\label{eq:V05 cases}
\end{align}

 \paragraph{$g=1$, $n=2$.} This is the first non-trivial case at higher genus and we obtain with the help of the intersection numbers on $\bM_{1,2}$ listed in Appendix~\ref{app:intersection numbers}
 \begin{align} 
\frac{V_{1,2,0}(\boldsymbol{\alpha})}{(2\pi^2)^2}&=\frac{1}{48}\left(\left(\alpha_1^2+\alpha_2^2-2\right)^2-\left(\theta(\alpha_1+\alpha_2-1)-1\right)^2\right) \\
&=\frac{1}{48}\begin{cases}
(1-\alpha_1^2-\alpha_2^2)(3-\alpha_1^2-\alpha_2^2)\ ,   &\alpha_1+\alpha_2<1\ , \\
\prod_i (1-\alpha_i)\left(1-\sum_i \alpha_i(\alpha_i-1)-\alpha_1\alpha_2\right)\ ,   &\alpha_1+\alpha_2\ge 1\ .
\end{cases}
 \end{align}
 One again checks that this is consistent with the string and dilaton equation, as well as with positivity of the volume.

\section{Two dimensional gravity and random matrices}\label{sec:JTRMT}
We begin with a brief review of the duality between pure JT gravity and matrix integrals, and the proposal of \cite{Turiaci:2020fjj} for the matrix integral computing the gravity path integral for deformations of JT gravity with a gas of defects. Then, we describe a way to perform the explicit gravitational path integrals using the mathematical tools developed in the previous section, and show in some simple examples it matches with the matrix integral.  

\subsection{JT gravity and matrix integrals}\label{sec:JTmi}
The following type of matrix integrals will be relevant 
\be 
    \mathcal{Z} = \int \d H  \ \e^{- L \Tr V(H)}\ ,
\ee
where $H$ is an $L\times L$ Hermitian matrix acting on an $L$-dimensional Hilbert space. The measure $\d H$ is a product over the real components of the matrix, and $V(H)$ is an arbitrary function: the matrix potential. In the large $L$ limit, keeping the matrix potential fixed, the Feynman diagrams of this matrix integral triangulate a two dimensional surface and the partition function has a topological expansion. 

We will need a generalization to surfaces with boundaries, and we consider expectation values of resolvant insertions defined by 
\be 
    R(E_1,\ldots,E_n) = \frac{1}{\mathcal{Z}}\int \d H \ \e^{- L \Tr V(H)} \Tr \frac{1}{E_1-H} \, \cdots\,  \Tr \frac{1}{E_n-H}\ ,
\ee
where the parameters $E_i$ for $i=1,\ldots, n$ can be complex numbers. This expression has contributions from disconnected diagrams and we define $R_{\rm conn.}(E_1,\ldots, E_n)$ to be the connected part of this correlator. In the large $L$ limit these correlators have a topological expansion alluded above
\be 
    R_{\rm conn.}(E_1,\ldots,E_n) \simeq \sum_{g=0}^{\infty} \frac{R_{g,n}(E_1,\ldots,E_n)}{L^{2g+n-2}}\ ,
\ee
where we emphasize that this expansion is only asymptotic. It is also convenient to define a 't Hooft expansion for the density of states $\rho(E) = \big\langle  \sum_{i=1}^L \delta(E-E_i)\big\rangle$. The sum is over the eigenvalues of a matrix realization $H$ and the bracket denotes a normalized expectation value over the matrix ensemble. This quantity also has an expansion in terms of $\rho(E) \simeq \sum_g L^{1-2g}\rho_{g}(E)$. Importantly, the leading order term $\rho_0(E)$ is uniquely specified in terms of the matrix potential, and therefore either quantity gives a specification of the matrix integral.   

The application to Weil-Petersson volumes, and to JT gravity, requires a further limit: the double-scaling limit. At finite but large $L$, the leading order density of states has a compact support and the normalization of $\rho_0$ is fixed through $\int \d E\,  \rho(E) = L$. In the double scaling limit we scale $L$ together with a fine-tuning in the matrix potential such that $\rho_0(E)$ has a non-compact support but remains finite. There is now a new topological expansion in terms of the overall scale of the leading order density of states, which we call $\e^{S_0}$, and is given by\footnote{In the rest of this paper all matrix integrals will be double-scaled and therefore we do not introduce new notation to distinguish $R_{g,n}(E_1,\ldots,E_n)$ and $\rho_g(E)$ from the analogous parameters in the 't Hooft limit.}
\be 
    R_{\rm conn.}(E_1,\ldots,E_n) \simeq \sum_{g=0}^{\infty} \frac{R_{g,n}(E_1,\ldots,E_n)}{(\e^{S_0})^{2g+n-2}}\ , \qquad \rho(E) \simeq \sum_{g=0}^{\infty} \frac{\rho_{g}(E)}{(\e^{S_0})^{2g-1}}\ .
\ee
This is a topological expansion in the new parameter $\e^{S_0}$. In this case the Feynman diagrams triangulating two dimensional surfaces become continuous. In the application to ${\rm NAdS}_2/{\rm NCFT}_1$ holography the matrix is interpreted as the disordered boundary hamiltonian $H$ of the quantum system dual to the black hole described by pure JT gravity \cite{Saad:2019lba}. 

The matrix model relevant to the Weil-Petersson volumes is a Hermitian ensemble in the double-scaling limit with the following leading order density of states 
\be \label{eq:JTdos}
    \rho_{\sf JT}(E) = \frac{\sinh(2\pi \sqrt{E})}{4\pi^2}\ ,\qquad E>0\ ,
\ee
and vanishes otherwise. At each order in the genus expansion, the WP volumes are related to the matrix model resolvant in the following way
\be \label{eq:MMV}
    V_{g,n}(b_1,\ldots,b_n) =(-1)^n \int_{\mathcal{C}} \, R_{g,n}(-z_1^2,\ldots,-z_n^2)\,  \prod_{j=1}^n \frac{\d z_j }{2\pi i}\,  \frac{2z_j}{b_j} \, \e^{b_j z_j}\ .
\ee
The contour $\mathcal{C}$ runs along the imaginary axis with a large enough real part appropriate for an inverse Laplace transform. It was shown by Eynard and Orantin \cite{Eynard:2007fi} that precisely the quantities $V_{g,n}(b_1,\ldots,b_n)$ constructed from this matrix integral reproduce the Weil-Petersson volumes of moduli space of hyperbolic surfaces with geodesic boundaries. This is proven in a straightforward way by matching the topological recursion relation that these volume satisfy, derived by Mirzakhani \cite{Mirzakhani:2006fta}, with the loop equations of the matrix integral in the double scaling limit. A nice presentation of these results can be found in \cite{Stanford:2019vob}.

The matrix model described above also computes the JT gravity path integral at each order in the genus expansion. First of all, the JT gravity partition function on the disk, coming from the Schwarzian reparametrization mode, reproduces the leading spectral density \eqref{eq:JTdos}. Another case that has to be treated separately is the genus zero contribution to the partition function with two boundaries, which again matches with the universal matrix model answer. Finally, other than those two cases, the partition function on any surface with any number of boundaries and handles can be written as 
\be \label{eq:SSSZgn}
    Z_{g,n}(\beta_1,\ldots,\beta_n) =\left[ \prod_{j=1}^n \int_0^{\infty} b_j\,  \d b_j\  Z_{\rm tr}(\beta_j,b_j)\right] V_{g,n}(b_1,\ldots,b_n)\ ,
\ee
where we defined the trumpet contribution as usual 
\begin{equation}
    Z_{\rm tr}(\beta,b) = \frac{\e^{-\frac{b^2}{4\beta}}}{2\sqrt{\pi\beta}}\ .
\end{equation}
In deriving this formula one uses the fact that to each of the $n$ boundaries there is a homotopically equivalent geodesic closest to it with length $b_j$. The path integral over the boundary graviton produces the factor of $Z_{\rm tr}(\beta,b)$ while the path integral in the interior produces with the WP volumes with geodesic boundaries, since the one-loop determinants in the bulk are trivial (we only consider orientable surfaces here). The measure $b \, \d b$ originates from the simple form of the Weil-Petersson volume form in Fenchel-Nielsen (length-twist) coordinates, see eq.~\eqref{eq:Weil-Petersson volume Fenchen-Nielsen coordinates}. 

As a simple example, we show the decomposition for $g=n=1$:
\begin{align} 
  Z_{1,1}(\beta)  &= \hspace{0.1cm}\begin{tikzpicture}[scale=0.3, baseline={([yshift=-0.1cm]current bounding box.center)}]
  \draw[thick] (-1,0) ellipse (0.25 and 1.8);
  \draw[thick] (-1,1.8) to [out=-40,in=180]  (.5,0.65) to [out=0,in=210] (2,1.5) to [out=30,in=90] (3.5,0) to [out=-90,in=-30] (2,-1.5) to [out=150,in=0] (.5,-0.65) to [out=180,in=40] (-1,-1.8);
  \draw[thick] (2,-0.9) to [bend right = 50] (2,0.9);
  \draw[thick] (2.2,-0.6) to [bend left = 50] (2.2,0.6);
    \end{tikzpicture}\\
    &=  \int_0^\infty b\,  \d b \, \beta\,\, \begin{tikzpicture}[scale=0.4, baseline={([yshift=-0.1cm]current bounding box.center)}]
    \draw[thick] (6.5,0) ellipse (0.3 and 1.5);
\draw[thick] (8.5,0) ellipse (0.1 and 0.7);
\draw[thick] (6.54,1.49) to [bend right=20] (8.5,0.7);
\draw[thick] (6.54,-1.49) to [bend left=20] (8.5,-0.7);
\end{tikzpicture}\,\, b\,\, \times\, \,b\,\, \begin{tikzpicture}[scale=0.5, baseline={([yshift=-0.1cm]current bounding box.center)}]
  \draw[thick] (0.5,0) ellipse (0.1 and 0.65);
  \draw[thick]  (.5,0.65) to [out=0,in=210] (2,1.5) to [out=30,in=90] (3.5,0) to [out=-90,in=-30] (2,-1.5) to [out=150,in=0] (.5,-0.65);
  \draw[thick] (2,-0.9) to [bend right = 50] (2,0.9);
  \draw[thick] (2.2,-0.6) to [bend left = 50] (2.2,0.6);
    \end{tikzpicture}\\
    &=  \int_0^\infty b \, \d b\ Z_{\rm tr}(\beta,b) \, V_{1,1}(b)\ .
\end{align}

Each boundary in the gravity path integral with Dirichlet boundary conditions corresponds in the matrix integral to an insertion of $\Tr(\e^{-\beta H})$. This is easily related to the resolvant by the integral transform $R(E) = - \int_0^{\infty} \d\beta \, \e^{\beta E} \Tr(\e^{-\beta H})$. Applying this integral transform to \eqref{eq:SSSZgn} and comparing with \eqref{eq:MMV} proves the equivalence of JT gravity with the matrix integral with leading density of states \eqref{eq:JTdos}. 

\subsection*{The string equation}
It will be useful, for the upcoming discussion, to introduce the string equation. A matrix integral is determined, up to a double-scaling limit, by a choice of a single function. This can be either the leading order density of states or the matrix potential. There is yet a third way to specify a matrix integral: through the string equation. It is determined through couplings $t_{k\ge 0}$ assembled in a function $\mathcal{F}(u) \equiv \sum_k t_k u^k$. The leading order density of states is related to the string equation by an integral transform 
\be \label{eq:rhofromF}
    \rho_0(E) = \frac{1}{2\pi} \int_{E_0}^E \frac{\d u}{\sqrt{E-u}} \, \partial_u \mathcal{F}(u)\ ,
\ee
where $E_0$ is the largest root of $\mathcal{F}(E_0)=0$. 

The string equation arises from the application of the orthogonal polynomial method to the matrix integral, combined with the double-scaling limit \cite{Douglas:1989ve, Gross:1989vs}. In this approach higher-genus corrections are computed by Taylor expanding $\mathcal{F}(u)$ in $u$, and replacing $u^k$ by the Gelfand-Dickii differential operators, which are given by an expansion in powers of $\e^{-S_0}$, with $u^k$ being their leading contribution. There is a machinery that turns this finite $S_0$ string equation into matrix model partition functions. We will not explain this procedure, since we will not use it in this article, but a review on this approach can be found in \cite{Johnson:2019eik}.

Combining \eqref{eq:rhofromF} and \eqref{eq:JTdos}, we conclude the matrix integral computing Weil-Petersson volumes has a string equation
\be 
    \mathcal{F}_{\sf JT}(u) = \frac{\sqrt{u}}{2\pi} \, I_1(2\pi\sqrt{u})\ .
\ee
We will find in the next sections that this is the simplest way to characterize the matrix integrals relevant for our purposes. Nevertheless, depending on the application, for example to compute higher genus corrections, it can be more convenient to extract the leading order spectral curve from \eqref{eq:rhofromF} and apply the loop equations.

\subsection{Deformations of JT gravity and matrix integrals}\label{sec:mmvcp}
In this section we present the matrix model that we conjecture computes the generating function of volumes of moduli space of surfaces with cone points, with the volume measure derived in section \ref{sec:WPform}. Equivalently, it is the matrix model computing the gravitational path integral of JT gravity deformed by a gas of generic defects.

Before doing so, it is useful to introduce the defect generating function, which depends on a complex variable $y$, and encodes the defects couplings $\lambda$ and deficit angles $\alpha$ as
\begin{equation}
    W(y) = \sum_i \lambda_i \, \e^{-2\pi(1-\alpha_i)y}\ .
\end{equation}
The sum can be over an arbitrary number of defects (although if we want to compute a volume with $n$ defects we need at least $n$ terms in the sum).

Lets begin by describing the sharp defect case, corresponding to all $0\leq \alpha_i \leq 1/2$. To compute the gravitational path integral of JT gravity deformed by this type of defects we need to sum over the number of handles, but also over the number of defects. Lets focus first on a single defect species. At fixed genus $g$ the partition function is given by 
\begin{equation}
Z_{g,n}(\beta_1,\ldots,\beta_n) = \sum_{k=0}^{\infty} \frac{\lambda^{k}}{k!} \left[\prod_{j=1}^n \int_{0}^{\infty} b_j \, \d b_j \ Z_{\rm tr}(\beta_j,b_j)\right] V_{g,k,n}(\underbrace{\alpha,\ldots,\alpha}_{k-{\rm terms}};b_1,\ldots,b_n)\ .
\end{equation}
The integer $k$ labels the number of defects. When the defect is sharp a geodesic homotopic to the boundary still exists and therefore we integrate over its length including the boundary graviton partition function $Z_{\rm tr}(\beta,b)$. The path integral in the interior is given by the WP volume with $k$ defects divided by $k!$ since in the gravitational path integral the defects are indistinguishable. Finally the factor of $\lambda^k$ gives the weight of each defect insertion. The generalization to an arbitrary number of defect species $s$ is straightfoward: the sum is over $s$ non-negative integers $\{k_1,\ldots, k_s\}$ labeling the number of defects of type $1$ to $s$. Each species comes with a factor of $\lambda_i^{k_i}/k_i!$ and the WP volume now has $k_i$ defects of type $i$.

From this point of view, the  function $W(y)$ is a particular way to package the information of the defect couplings and angles. When the addition of the gas of defects is modeled by a change in the dilaton potential, the shift in $U(\phi)$ is related to $W(y)$ identifying $y$ with the dilaton. See section \ref{sec:con}, and also \cite{Turiaci:2020fjj} for a discussion on this identification.

The case with sharp defects simplified the gravitational path integral in two ways. First, it allows to write the answer in terms of WP volumes with geodesic boundaries. When defects are blunt, its not always true that a geodesic exists separating the defect from the NAdS  boundary. Second, the WP volumes with sharp cone points are given by \cite{Tan:2006,Do_cone}:
\begin{equation}\label{eq:WPsharpd}
V_{g,k,n}(\alpha_1,\ldots,\alpha_{k};b_1,\ldots,b_n)=V_{g,n+k}(2\pi \i\alpha_1 ,\ldots, 2\pi {\rm i} \alpha_k; b_1,\ldots,b_n).
\end{equation}
This is again not true for blunt defects, and an improved definition of these volumes was given in section \ref{sec:Weil Petersson form}. When these two simplifications occur it is possible to prove an equivalence between deformations of JT gravity by sharp defects and a double-scaled matrix integral with (tree level) string equation
\be \label{eq:sharpse}
    \mathcal{F}_{\sf dJT}(u) = \frac{\sqrt{u}}{2\pi} I_1(2\pi\sqrt{u}) + \sum_{i=1}^s \lambda_i I_0 (2\pi \alpha_i \sqrt{u})\ , \qquad ({\rm for}\, 0\leq \alpha_i\leq 1/2)\ .
\ee
From this equation we can obtain the location of the edge of the spectrum by finding the largest root of $\mathcal{F}_{\sf dJT}(E_0)=0$ and then use \eqref{eq:rhofromF} to obtain the density of states.\footnote{Its not possible to find $E_0$ analytically but one can compute it in perturbation theory around $\lambda=0$, or numerically for arbitrary deformations.} The string equation can be rewritten in the following involving $W(y)$, instead of the defect weights and angles separately:
\be \label{eq:SEGDJT}
   \mathcal{F}_{\sf dJT}(u) = \int_{\mathcal{C}} \frac{\d y}{2\pi {\rm i}} \e^{2\pi y} \left( y-\sqrt{y^2-u-2W(y)}\right)\ .
\ee
The contour is again along the imaginary axis with large enough real part, such that singularities are to the left. Taylor expanding in $\lambda$ we can see that the zeroth order term matches with the JT string equation, the linear terms match with \eqref{eq:sharpse} and any higher order term in $\lambda$ vanish if the defects are sharp. This leads to the following expression for the density of states
\be \label{eq:DOSGDJT}
    \rho_{\sf dJT}(E)= \frac{1}{2\pi} \int_{\mathcal{C}} \frac{\d y}{2\pi {\rm i}} \e^{2\pi y} \tanh^{-1} \left(\sqrt{\frac{E-E_0}{y^2-2W(y)-E_0}}\right)\ ,
\ee
The edge of the spectrum $E_0$ depends non-trivially on the defect couplings so it is important to include it as a parameter. With this information we can compute any observable by either using the matrix model loop equations, or equivalently the string equation. 

So far we described a theory of gravity we can solve. What happens when we try to extend the solution to deformations of JT gravity by generic defects with $0\leq \alpha \leq 1$? \emph{The proposal of \cite{Turiaci:2020fjj}, motivated by studying the minimal string theory, is that the matrix model with  string equation given by \eqref{eq:SEGDJT} is the correct continuation to a gas of defects with arbitrary deficit angles}. 

When the defects are blunt $1/2<\alpha<1$ the tree level string equation \eqref{eq:SEGDJT} is \textbf{not} equal to one derived for sharp defects \eqref{eq:sharpse}, since now higher powers of $\lambda$ are non-zero, therefore using \eqref{eq:sharpse} in this regime gives the wrong answer. For example, if we used \eqref{eq:sharpse} in the blunt defect regime we could get negative answers for the partition function, analogous to the discussion around \eqref{eq:four punctured sphere with defects wrong answer}, which would not make sense.

To prove the conjecture of \cite{Turiaci:2020fjj} we would need to provide a gravitational calculation of the partition function, but we run into two troubles already mentioned above. First, if a geodesic exists in a hyperbolic metric separating the boundaries from the defects and handles, we cannot use Mirzakhani's formula \eqref{eq:WPsharpd} anymore since the Weil-Petersson measure is not correct. We will show in the next section how to solve this issue: the answer from the matrix model with tree level string equation \eqref{eq:SEGDJT} matches the new WP volumes we defined in section \ref{sec:Weil Petersson form}. Second, there are geometries with multiple defects that do not have geodesics anywhere. Then the procedure of SSS of separating into trumpets and interior surfaces bounded by geodesics does not work. This can be solved using the techniques of equivariant localization developed in \cite{Stanford:2017thb, Eberhardt:2022wlc}: The configurations with no geodesics only take places without handles and the path integral localizes into configurations where all the defects merged into one. The answer reduces to the single defect on the disk with a small modification of the one-loop determinant. We will show this is also consistent with the results coming from \eqref{eq:SEGDJT}.

\subsection{Dilaton gravity path integral evaluation}

In this section we will explain how to perform the gravitational path integral for deformations of JT gravity by a gas of generic defects, not necessarily sharp. We will first work out a simple example and then outline the general procedure. 

\subsubsection*{An example}
As a simple example to illustrate the idea we consider the case of JT gravity deformed by a single defect species with weight $\lambda$ and opening angle $2\pi \alpha$. We will first write down the matrix model result and then present the gravity interpretation. 

First we compute the partition function $Z_{\sf dJT}(\beta)$ at genus zero, perturbatively in $\lambda$. The tree-level string equation has the following form:
\begin{align} 
    \mathcal{F}_{\sf dJT}(u) &=\int_{\mathcal{C}} \frac{\d y}{2\pi {\rm i}}\ \e^{2\pi y} \left( y-\sqrt{y^2-u-2\lambda \e^{-2\pi(1-\alpha)y}}\right),\nonumber\\
    &=\sum_{L=0}^{\lfloor\frac{1}{1-\alpha}\rfloor} \frac{\lambda^L}{L!} \left( \frac{2\pi(1-L(1-\alpha))}{\sqrt{u}}\right)^{L-1} I_{L-1}(2\pi(1-L(1-\alpha))\sqrt{u})\ . 
\end{align}
The term with $L=0$ is pure JT gravity and the term with $L=1$ matches the sharp defect answer. The higher order terms in $\lambda$ are new. Order by order in $\lambda$, we can first compute $E_0$ and then evaluate the leading order $\rho_{\sf dJT}(E)$ using \eqref{eq:rhofromF}. Using this density of states we can compute  $Z_{\sf dJT}(\beta) = \int_{E_0}^E \d E \, \e^{-\beta E} \rho_{\sf dJT}(E)$, the partition function.

\paragraph{Sharp defect} Before presenting the result for $\alpha>1/2$ let us recall the answer for sharp defect $0\leq \alpha \leq 1/2$. In this case the string equation is 
\begin{equation}
 \mathcal{F}_{\sf dJT}(u) = \frac{\sqrt{u}}{2\pi} I_1(2\pi\sqrt{u}) + \lambda I_0 (2\pi \alpha \sqrt{u})\ .
\end{equation}
and the edge of the spectrum is located perturbatively at 
\begin{equation}
  E_0 = -2\lambda -2\pi^2(1-2\alpha^2) \lambda^2 + \frac{2\pi^4}{3}(15 \alpha^4 - 18 \alpha^2 +5) \lambda^3 + \mathcal{O}(\lambda^4)\ .
\end{equation}
The tree-level partition function extracted from this string equation is:
\begin{multline}
    \e^{-S_0} Z_{\sf dJT}(\beta) = \frac{\e^{\frac{\pi^2}{\beta}}}{4\sqrt{\pi}\beta^{3/2}} + \lambda \frac{\e^{\frac{\pi^2\alpha^2}{\beta}}}{2\sqrt{\pi \beta}}\\
    + \frac{\lambda^2}{2!} \frac{\sqrt{\beta}}{\sqrt{\pi}} + \frac{\lambda^3}{3!} \frac{2\sqrt{\beta}(\pi^2(1-3\alpha^2)+\beta)}{\sqrt{\pi}}+\mathcal{O}(\lambda^4)\ .
\end{multline}
The first term, of order $\lambda^0$, is the pure JT gravity partition function on the disk. The term of order $\lambda$ is the partition function of JT gravity on the hyperbolic disk with a single defect in the bulk, computed in \cite{Mertens:2019tcm}. The second and third terms are given by the integral of the WP volume with $b\to 2\pi \i \alpha$ replacement for defects:
\begin{align}
    \frac{\sqrt{\beta}}{\sqrt{\pi}} &= \int_0^\infty b\, \d b\, Z_{\rm tr}(\beta,b) \, V_{0,3}(2\pi \i \alpha, 2\pi \i \alpha,b)\ ,\\
  \frac{2\sqrt{\beta}\big(\pi^2(1-3\alpha^2)+\beta\big)}{\sqrt{\pi}}  &=\int_0^\infty b\, \d b\, Z_{\rm tr}(\beta,b) \, V_{0,4}(2\pi \i \alpha, 2\pi \i \alpha,2\pi \i \alpha,b)\ ,
\end{align}
and similarly for higher orders in $\lambda$. Of course is this guaranteed since one can actually prove this works to all orders in both genus and defect expansion, see  \cite{Maxfield:2020ale}.

\paragraph{Defect with $\frac{1}{2}<\alpha\leq \frac{2}{3}$} This range is chosen such that the string equation is supplemented by only a quadratic term in $\lambda$. More explicitly, 
\begin{equation}
 \mathcal{F}_{\sf dJT}(u) = \frac{\sqrt{u}}{2\pi} I_1(2\pi\sqrt{u}) + \lambda I_0 (2\pi \alpha \sqrt{u}) +    \frac{\lambda^2}{2} \frac{2\pi(2\alpha-1)}{\sqrt{u}} I_{1}(2\pi(2\alpha-1)\sqrt{u})\ .
\end{equation}
The position of the edge of the spectrum is located (perturbatively) at 
\begin{equation}
    E_0 = -2\lambda -4\pi^2(1-\alpha)^2 \lambda^2 + \frac{2\pi^4}{3}(57 \alpha^4 - 120 \alpha^3 + 72\alpha^2 -8) \lambda^3 + \mathcal{O}(\lambda^4)\ . 
\end{equation}
For $\alpha>2/3$ we can have cubic or higher orders as well and we have jumps in the highest power for all $\alpha = (n-1)/n$ with integer $n$. The matrix integral prediction from the proposed string equation is now different, given by\footnote{Notice the right hand side is different than the one in equation (4.17) of \cite{Turiaci:2020fjj}, which had a mistake in the calculation.}
\begin{multline}\label{Zbdck}
    \e^{-S_0} Z_{\sf dJT}(\beta) = \frac{\e^{\frac{\pi^2}{\beta}}}{4\sqrt{\pi}\beta^{3/2}} + \lambda \, \frac{\e^{\frac{\pi^2\alpha^2}{\beta}}}{2\sqrt{\pi \beta}} \\
    + \frac{\lambda^2}{2!} \, \frac{\sqrt{\beta}}{\sqrt{\pi}}\e^{\frac{\pi^2(1-2\alpha)^2}{\beta}} + \frac{\lambda^3}{3!} \, \frac{2\sqrt{\beta}(\pi^2(2-3\alpha)^2+\beta)}{\sqrt{\pi}}+ \mathcal{O}(\lambda^4)\ .
\end{multline}
The two terms in the first line are the same as in the sharp defect case. The first term in the second line corresponds to two defects, but if $\alpha>1/2$ there is no geodesic separating them from the NAdS boundary. We will explain below how to reproduce the result using equivariant localization in the next section. The last term corresponds to three defects in the bulk, but now as long as $\alpha<2/3$ there will be a geodesic separating them from the NAdS boundary. Therefore we should be able to apply the gluing procedure of SSS
\be 
\int_0^{\infty} b\, \d b \ Z_{\rm tr}(\beta,b) \, V_{0,3,1}(\alpha,\alpha,\alpha;b) \overset{?}{=}\frac{2\sqrt{\beta}\big(\pi^2(2-3\alpha)^2+\beta\big)}{\sqrt{\pi}}\ .
\ee
This result can be reproduced by a WP volume given by 
\be \label{eq:V04MMP}
    V_{0,3,1}(\alpha,\alpha,\alpha;b) = 2\pi^2 \Big( \Big( \frac{b}{2\pi}\Big)^2 + (2-3\alpha)^2\Big).
\ee
We actually already computed this in section \ref{sec:Weil Petersson form} using the volume form we derived. In equation \eqref{eq:V04 cases} we present the WP volume with four defect. In our case, we have three blunt defects and a geodesic boundary. But this is equal to the fourth case in \eqref{eq:V04 cases} where $\alpha_2=\alpha_3=\alpha_4>1/2$ and after continuing $\alpha_1\to \frac{b}{2\pi \i}$, then \eqref{eq:V04 cases} matches with \eqref{eq:V04MMP}. Therefore we reproduced the order $\lambda^3$ term from a gravity calculation.

We can consider the evaluation of matrix model observables,  that involve in the gravity description a defect partition function over geometries with no geodesics. We already saw one example above, the quadratic term in $\lambda$ in equation \eqref{Zbdck}. These are easy to compute exactly from the matrix integral. First, this situation only takes place at genus zero and a single boundary, and therefore we do not need to use the loop equations. This is a consequence of Gauss-Bonnet theorem. Second, they come with exponential factors making it easy to identify their origin, see Appendix D of \cite{Turiaci:2020fjj}. Their contribution to the partition function is
\be \label{eq:mergedeffrommm}
     Z_{\sf dJT}(\beta) =\e^{S_0} \sum_{k=0}^{\lfloor \frac{1}{1-\alpha} \rfloor} \frac{\lambda^k}{k!} (2\beta)^{k-1} \frac{\e^{\frac{\pi^2(1-k(1-\alpha))^2}{\beta}}}{2\sqrt{\pi \beta}} + \ldots\ ,
\ee
where the dots denote all terms that are either higher order in $\lambda$ or $\e^{-S_0}$ and therefore can be computed using the WP volumes in section \ref{sec:Weil Petersson form} (for example, they are polynomials in $\sqrt{\beta}$). The exponential term looks exactly like the one corresponding to the hyperbolic disk with a single defect with an effective opening angle $2\pi \alpha_{\rm eff} =2\pi( 1-k(1-\alpha))$, which is precisely the angle that would result from a merger of $k$ defects. We explain below how to reproduce this terms exactly from a gravitational path integral argument.

\subsubsection*{General case with geodesic boundaries}
To summarize, except for a few cases at genus zero described in the previous paragraph, the defects and handles in gravity are separated from the NAdS boundaries by geodesics. Then the gravity partition function is given by an expression looking similar to the sharp defect answer, which for simplicity we write in the case of a single defect species 
\begin{equation}\label{eq:gendzgngn}
Z_{g,n}(\beta_1,\ldots,\beta_n) = \sum_{k=0}^{\infty} \frac{\lambda^{k}}{k!} \left[ \prod_{j=1}^n\int_{0}^{\infty} b_j \d b_j Z_{\rm tr}(\beta_j,b_j)\right] V_{g,k,n}(\underbrace{\alpha,\ldots,\alpha}_{k\text{ terms}};b_1,\ldots,b_n)\ ,
\end{equation}
with a trivial generalization to multiple species. The difference is that now $V_{g,k,n}$ is the WP volume with $k$ defects and $n$ boundaries, computed using the WP measure we derive in section \ref{sec:Weil Petersson form}. It is not the analytic continuation of the expression with $n+k$ boundaries. 

We leave as an open interesting mathematical problem to find a proof that the volumes we defined in section \ref{sec:Weil Petersson form} match the matrix integral prediction. In this article we simply verify it in some concrete cases.

\subsubsection*{Cases without geodesic boundaries}
The equation \eqref{eq:gendzgngn} covers all cases, except for the genus zero disk with a number of defects such that $\sum_i(1-\alpha_i) <1$. This is precisely the case where no geodesic is present in the geometry. 
We shall now explain how to compute \eqref{eq:mergedeffrommm} from a bulk calculation. This can be done using a version of equivariant localization (a.k.a. Duisterman-Heckman formula in the symplectic context). The method was explained in \cite{Stanford:2017thb} in the case of the disk without defects and generalized to arbitrary surfaces with geodesic boundaries in \cite{Eberhardt:2022wlc}. The case explained here has not appeared before in the literature, but is actually one of the simplest cases, since the integral localizes to a single point. For that reason, we will be brief in our discussion and refer to \cite{Stanford:2017thb, Eberhardt:2022wlc} for more details.

Let us denote by $\mathscr{M}_{0,k}$ the universal moduli space of the disk with $k$ marked points including fluctuations of the Schwarzian mode. This includes the simple cases of the universal Teichm\"uller space $\mathscr{M}_{0,0}=\Diff(\text{S}^1)/\PSL(2,\RR)$ and a generic Virasoro coadjoint orbit  $\mathscr{M}_{0,1}=\Diff(\text{S}^1)/\U(1)$ \cite{Witten:1987ty}. For $k \ge 2$, the moduli space requires an appropriate compactification as in the case without boundaries that adds boundary divisors corresponding to colliding marked points. We denote the compactified moduli spaces by $\overline{\mathscr{M}}_{0,k}$.
These moduli spaces are infinite-dimensional and computing their volume directly does not make sense. It does make sense however to compute their equivariant volume, which is a fancy mathematical way to include the temperature $\beta$ of the boundary into the computation. It is related to the chemical potential for the $\U(1)$ action on $\mathscr{M}_{0,k}$ that rotates the disk.

\begin{figure}
    \centering
    \begin{tikzpicture}[scale=.7]
        \draw[very thick] (0,0) circle (4);
        \draw[smooth, thick, samples=100,domain=0:360, red] plot({3.4*cos(\x)*(1+0.05*sin(19*\x)+0.05*cos(13*\x))}, {3.4*sin(\x)*(1+0.05*sin(19*\x)+0.05*cos(13*\x))});
        \fill[very thick] (1.3,.4) circle (.1) node[above] {1};
        \fill[very thick] (-.7,-1.4) circle (.1) node[above] {2};
        \fill[very thick] (-1.7,-.8) circle (.1) node[above] {3};
        \fill[very thick] (.3,.6) circle (.1) node[above] {4};
        \draw[very thick,->] (-30:4.3) arc  (-30:30:4.3);
        \node at (5.1,0) {$\U(1)$};
    \end{tikzpicture}
    \caption{The disk moduli space with $n$ marked points (here $n=4$). The surface is unchanged under $\U(1)$ rotations when all points coincide at the center of the disk and the asymptotic cutoff for the Schwarzian mode is round.}
    \label{fig:U1 fixed point}
\end{figure}
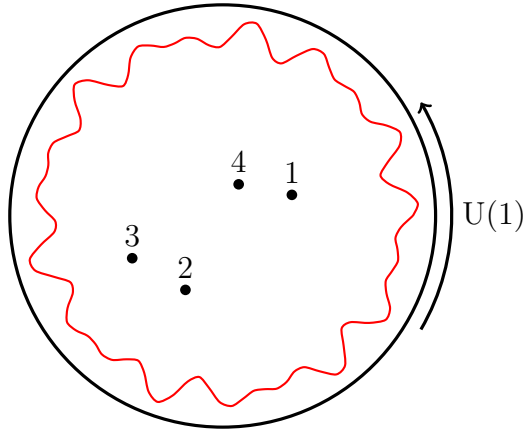

The equivariant localization formula states that
\be 
\int_{\overline{\mathscr{M}}_{0,k}} [\e^{\omega_\text{WP}}]_x=\frac{[\e^{\omega_\text{WP}}]_{x,0}(\star)}{\prod_j (xm_j)}\ . \label{eq:equivariant localization}
\ee
Here, $[\e^{\omega_\text{WP}}]_x$ is the equivariant completion of the Weil-Petersson measure on moduli space that we want to integrate. $x$ is the equivariant parameter, which is related to the boundary temperature as $x=\frac{1}{2\beta}$.\footnote{The proportionality constant in this relation is conventional and is given by the constant $\frac{\gamma}{\alpha}$ in the notation of \cite{Saad:2019lba}.}
There is a unique fixed point of the $\U(1)$-action corresponding to a surface where all conical defects coincide at the center of the disk, see Figure~\ref{fig:U1 fixed point}, since by assumption they are sufficiently blunt to be able to collide.
On the right hand side of \eqref{eq:equivariant localization}, we need to evaluate the equivariant Weil-Petersson measure at the fixed point $\star$.\footnote{In mathematical terms, this is the pullback of $[\e^{\omega_\text{WP}}]_x$ under the inclusion map $\iota: \star \longrightarrow \overline{\mathscr{M}}_{0,k}$.} In physical terms, this is the tree-level contribution to the `path-integral' on the left hand side. The product $\prod_j (x m_j)=\prod_j \frac{m_j}{2\beta}$ is the one-loop piece. Here, $m_j$ are the charges of the $\U(1)$ action on the tangent space of the fixed point. The tangent space breaks up into two parts: tangent directions for the Schwarzian mode and directions that move around the conical defects in the disk. The former leads to the charges $2,3,4,\dots$, while the latter lead to the charges $1,1,\dots,1$ ($k$ times).  The result is one-loop exact as was first shown for the disk partition function in \cite{Stanford:2017thb}.

Thanks to the pull-back property \eqref{eq:pullback Weil Petersson form separating divisor}, the on-shell action on the right hand side of \eqref{eq:equivariant localization} looks like the on-shell action for a single conical defect with defect angle $2\pi \sum_i (1-\alpha_i)$. 
We have then $[\omega_\text{WP}]_{\beta,0}=\frac{\pi^2}{\beta}(1-\sum_i (1-\alpha_i))^2$.\footnote{
This can be shown mathematically by using the restrictions (for $\star \in \overline{\mathscr{M}}_{0,1} \Diff(\text{S}^1)/\text{S}^1$) that $[\kappa_1]_{x,0}=0$ and $[\psi_1]_{x,0}=-x=-\frac{1}{2\beta}$ which is explained in \cite{Eberhardt:2022wlc}. Plugging in \eqref{eq:Weil-Petersson form with boundary classes} gives $[\omega_\text{WP}]_{x,0}=\frac{\pi^2\alpha^2}{\beta}$ for a single defect.} Putting the ingredients together leads to the following Weil-Petersson volume:
\be 
\int_{\overline{\mathscr{M}}_{0,k}} [\e^{\omega_\text{WP}}]_x=(2\beta)^{k-1}\frac{\e^{\frac{\pi^2}{\beta}(1-\sum_i (1-\alpha_i))^2}}{\prod_{n=1}^\infty (\frac{n}{2\beta})}=(2\beta)^{k-1}\frac{\e^{\frac{\pi^2}{\beta}(1-\sum_i (1-\alpha_i))^2}}{2\sqrt{\pi \beta}}\ ,
\ee
where we used zeta-function regularization for the infinite product. This matches with the matrix model computation \eqref{eq:mergedeffrommm}. In \eqref{eq:mergedeffrommm} we computed the generating function of defects which is responsible for the extra $\frac{1}{k!}$ since defects are treated as indistinguishable.

\subsection{More checks and the minimal string}
In the previous section we provides some checks involving the WP volume with four or five defects. We only required to use some specific cases out of all the possibilities considered in section \ref{sec:Weil Petersson form}. In this section we will compute, using the matrix integral, the sphere partition function with four and five defects and compare the WP volume computed in equations \eqref{eq:V04 cases} and \eqref{eq:V05 cases}, finding again a perfect match. 

The matrix model proposed in \cite{Turiaci:2020fjj} is the large $p$ limit of the matrix model dual to deformations of the $(2,2p+1)$ minimal string. Since the sphere partition function with defects was already computed in the past at finite $p$ we will review the relation to the minimal string and then take the large $p$ limit of the final answer. On the way, we will clarify the following point: the minimal string has two distinct sectors in the sphere, the odd and the even, depending on the nature of the minimal model sector of the operator insertions. We find this to be a subtle issue and only the large $p$ limit of the odd sector has a geometric interpretation. This issue does not arise for observables with at least one boundary. 

\subsubsection{The \texorpdfstring{$(2,2p+1)$}{(2,2p+1)} minimal string}
Before considering the proposed matrix model to compute the volumes of moduli space with defects, we will introduce the minimal string, which was the motivation of \cite{Turiaci:2020fjj} for the proposal. 

The minimal string is a theory of two dimensional gravity coupled to a minimal model CFT. We will be particularly interested in the coupling to the non-unitary series $(2,2p+1)$, labeled by the positive integer $p$. This theory of two dimensional gravity is conjectured to be dual to a one-matrix integral \cite{Kazakov:1989bc,Staudacher:1989fy}, which was made more precise in \cite{Moore:1991ir} and \cite{Belavin:2008kv}. 

When working in conformal gauge, one can rewrite the minimal string as a theory of Liouville gravity coupled to the minimal model. We can then turn the evaluation of the partition function on arbitrary surfaces of a given genus (with or without boundaries) to an integral over moduli space of the product of CFT correlators. This non-trivial calculation can be reproduced by the genus expansion of the double-scaled one-matrix integral with KdV couplings extracted from the following string equation
\be \label{eq:MSSE}
    \mathcal{F}_{\sf MS}(u) = \frac{2p+1}{(4\pi)^2}\left[ P_{p+1}(u_{\sf MS} )-P_{p-1}(u_{\sf MS} ) \right]\ , \qquad u_{\sf MS} \equiv 1+\frac{8\pi^2u}{(2p+1)^2}\ .
\ee
$P_{k}(x)$ denotes the Legendre polynomial of degree $k$. In writing this expression we have shifted and rescaled $u$ to remove the dependence on the bulk cosmological constant. See \cite{Turiaci:2020fjj} for the relation to more standard conventions in the minimal string literature. If boundaries are present, the insertion of $\Tr \left( \e^{-\beta H } \right)$ corresponds to fixed-length boundary conditions, see \cite{Mertens:2020hbs} for an account. Using \eqref{eq:rhofromF} we can obtain the leading order density of states of the matrix integral which is useful if one wishes to use the topological recursion 
\be 
    \rho_{\sf MS}(E) = \frac{1}{4\pi^2} \sinh\left( \frac{2p+1}{2} \cosh^{-1}\left(1+\frac{8\pi^2 E}{(2p+1)^2} \right)\right)\ .
\ee
Here we can see the advantage of our conventions, when $p\to\infty$ we recover the JT gravity density of states computing the WP volumes of moduli space of smooth hyperbolic surfaces \cite{Saad:2019lba}. To derive some of these formulas, the identities presented in Appendix A of \cite{Turiaci:2020fjj} are useful.

So far we reviewed the status of the matrix integral computing the volume of moduli space of smooth surfaces (related to JT gravity) and its finite $p$ generalization (the minimal string).\footnote{It is an interesting question to figure out whether the finite $p$ answer has an interpretation as an integral over moduli space. We will not attempt to do this in this paper and mostly consider the $p\to\infty$ limit.} But what about the matrix integral computing the volume of surfaces with cone points? This will be related to deformations of the minimal string where we shift the action by a minimal string tachyon operator. 

The $(2,2p+1)$ minimal model has a family of primary operators $\Phi_k$ labeled by an integer $k=0,\ldots, p-1$ with $k=0$ corresponding to the identity operator and $k=p-1$ having the lowest dimension. The minimal string ``tachyon'' operators $\mathcal{O}_k$ are labeled by the same integer. They consist of an integrated minimal model primary multipled, in the conformal gauge, by a Liouville primary $\e^{2\alpha_k \phi}$ with parameter $\alpha_k$ such that the operator is marginal. The deformed minimal string consist of a gravity theory with action
\be
    I =I_{\rm MS} + \sum_{k=0}^{p-1} \lambda_k \mathcal{O}_k\ ,\qquad \mathcal{O}_k = \int \Phi_k \e^{2\alpha_k \phi} \ ,
\ee
where $I$ is the total Euclidean action, $I_{\rm MS}$ is the sum of Liouville and minimal model actions and $\lambda_k$ are the couplings of the deformation. In particular the case $k=0$ corresponds to a shift of the bulk cosmological constant. We will assume this theory makes sense to any order in perturbation theory in $\lambda$. We will not address whether this theory makes sense non-perturbatively in the deformation parameters. 

In the next section we will review the relation between insertions of the tachyon operators $\mathcal{O}_k$ and the presence of cone points in the large $p$ limit.

The main result of \cite{Belavin:2008kv} is bootstrap of the possible string equation of the deformed minimal string. The sphere correlators derived from it have to be consistent with the minimal model fusion rules. This was a question raised and partially solved in \cite{Moore:1991ir}. The answer is given by the following tree level string equation as a function of deformation couplings
\be 
    \mathcal{F}_{\sf dMS} (u) = -\frac{(p+\frac{1}{2})^2}{4\pi^2}\oint \frac{\d z}{2\pi {\rm i}} \frac{\big(1-2 u_{\sf MS}z + z^2 - \frac{4\pi^2}{(p+\frac{1}{2})^2}\sum_{k=1}^{p-1} 2 \lambda_k z^{k+2} \big)^{\frac{1}{2}}}{z^{p+2}}\ .
\ee
The overall normalization is set to match with \eqref{eq:MSSE} when the deformation parameters are set to zero. This equation is the finite $p$ generalization of \eqref{eq:SEGDJT}. At large $p$, the parameter $k$ of the tachyon is related to the deficit angle by
\be
    k = p (1-\alpha)\ , \qquad p\to\infty\ .
\ee
The normalization of the tachyon coupling is chosen to match with the defect coupling at large $p$. The deformation by the identity $k=0$ corresponds to $\alpha=1$, a gas of marked points. The analog of sharp defects corresponds to operators with $k > \frac{p}{2}$, where the contour integral becomes linear in the $\lambda$'s.

\paragraph{Even vs odd sector in the matrix integral:} On the sphere, correlation numbers between tachyon operators with label $k_1,\ldots,k_n$ behave very differently depending on whether $k_1+\ldots+k_n$ is odd or even. This raises the question of whether the results in the large $p$ limit depend on whether the limit on $k$ is taken in the even or odd sector. This is a disctinction not raised before in the context of relating the minimal string with JT gravity. We will see in the next section that the large $p$ limit differs in the odd and even sector. Moreover, only the odd sector has a geometric interpretation connecting the minimal string with JT gravity.

In the rest of this section we will study some specific non-trivial examples that support our conjecture that the volumes defined in section \ref{sec:WPform} match with the ones derived from the matrix integral of section \ref{sec:mmvcp}. More specifically, we will reproduce the expressions \eqref{eq:V04 cases} and \eqref{eq:V05 cases} from the matrix integral.

\subsubsection{Four defects in a sphere}
The first case we will match is the volume of a surfaces of genus $g=0$ with four cone points. From the matrix integral point of view this is a subtle case: the genus zero partition function is not well-defined in the double-scaling limit and the result is ambiguous. Nevertheless, when taking derivatives with respect to $\lambda$ to compute the amplitude with four defects this ambiguity disappears and the result can be compared with the volume given in \eqref{eq:V04 cases}.

Fortunately, the sphere partition function with four insertions was already computed using the minimal string matrix integral, at finite $p$, in \cite{Belavin:2008kv} (see also \cite{ Tarnopolsky:2009ec,Artemev:2022hvu,Artemev:2022rng}). Since the matrix integral of section \ref{sec:mmvcp} is nothing else than the large $p$ limit of the minimal string matrix integral, the calculation is essentially the same.

The expression at finite $p$ depends on four tachyon operators labeled by $k_1$, $k_2$, $k_3$ and $k_4$. Without loss of generality we assume they are ordered according to $0\leq k_4\leq k_3\leq k_2\leq k_1 \leq p-1$. 
\paragraph{The odd sector:} The sphere four-point function in the odd sector is given by \cite{Belavin:2008kv}
\begin{equation}
V_{0,4,0}^p(\boldsymbol{\alpha}) =  2\pi^2\bigg( F(-2) - \sum_{i=1}^4 F(k_i-1) + F(k_{43|21})+F(k_{42|31})+F(k_{14|23})\bigg)\ ,
\end{equation}
where we use the index $p$ to emphasize this is computed with the finite $p$ matrix integral. Following \cite{Artemev:2022hvu} we introduced $k_{ij|lm} = {\rm min}(k_{ij},k_{lm})$ and $k_{ij}=k_i+k_j$. The function appearing in the right hand side is defined as
\be 
    F(k) = \frac{(p-k-1)(p-k-2)}{p^2}\, \Theta(p-2-k)\ .
\ee
We should emphasize that the normalization of this formula is different than the one used in the minimal string literature. We used the conventions introduced in \cite{Turiaci:2020fjj} that guarantees that the large $p$ limit matches with the defect result including the normalization. 

In the large $p$ limit, the deficit angles and the tachyon label are related by $k=p(1-\alpha)$. It is useful to note that in this limit the function becomes
\be 
   \lim_{p\to\infty}  F(p-p\alpha) = \alpha^2 \Theta(\alpha) = \theta(\alpha)\ , \qquad \lim_{p\to\infty}  F(x) =1\ ,
\ee
where in the second equality of the first equation we used the notation in \eqref{eq:V04 cases}, and in the second equation we take $x$ any number that does not scale with $p$. Using this limits its easy now to write down the large $p$ prediction:
\begin{equation}
\lim_{p\to\infty}\frac{V_{0,4,0}^p(\boldsymbol{\alpha})}{2\pi^2} =  1 - \sum_{i=1}^4 \alpha_i^2 + \sum_{i\leq i < j \leq 4}\theta(\alpha_i+\alpha_j-1) = \frac{V_{0,4,0}(\boldsymbol{\alpha})}{2\pi^2}\ ,
\end{equation}
where $V_{0,4,0}$ denotes the expression found in \eqref{eq:V04 cases}. This verifies our conjecture. 

When all defects are sharp the third term vanishes since for any pair of defects $\alpha_i +\alpha_j <1$. The fact that this case matches with the sharp defect WP volume was verified in \cite{Artemev:2022hvu} but follows more generally from the derivation in \cite{Maxfield:2020ale}. Here we verified the match also for the terms that are important when defects are blunt. 

The minimal string expression with four insertions was also reproduced from the continuum limit integrating the product of the minimal model and Liouville four-point function over moduli space \cite{Belavin:2006ex}.

\paragraph{The even sector:} We verified our conjecture for the odd sector of the minimal string. What about the even sector? We can compute the answer explicitly for the sphere with four defects from the minimal string. This is (up to a different choice of normalization) the large $p$ limit of equation (2.20) in \cite{Tarnopolsky:2009ec}. We find the following result:
\be 
    \lim_{p\to\infty}\frac{V_{0,4,0}^{{\rm even},p}(\boldsymbol{\alpha})}{2\pi^2}=\begin{cases}
    1-\alpha_1^2-\alpha_2^2-\alpha_3^2-\alpha_4^2\ , & \alpha_{34} < 1\ , \\
    2(1-\alpha_3)(1-\alpha_4)-\alpha_1^2-\alpha_2^2\ , & \alpha_{34} \geq 1,\, \alpha_{42} \le 1\ ,  \\
    (1-\alpha_4) (3-2 \alpha_{23}-\alpha_4)-\alpha_1^2\ ,\!\! & \alpha_{42}\geq 1,\, \alpha_{14} \le 1,\, \alpha_{23} \le 1\ , \\
        -\alpha_1^2+(2-\alpha_{234})^2\ , & \alpha_{23}\geq 1,\, \alpha_{14}\leq\alpha_{23},\\
  2(1-\alpha_4)(2-\alpha_{1234})\ , & \alpha_{14}\geq 1,\, \alpha_{14}>\alpha_{23}  ,\\
    0\ , & -\alpha_1+\alpha_{234}\geq 2\ ,
    \end{cases} 
\ee
where to simplify the formula we defined $\alpha_{ij} = \alpha_i + \alpha_j$ and $\alpha_{234} = \alpha_2 + \alpha_3 +\alpha_4$. Whenever the volume in the even and odd sectors are simultaneously non-vanishing, they match. But in the even sector, the volumes automatically vanish when $-\alpha_1 + \alpha_{234} \geq 2$ and can be non-zero when the Gauss-Bonnet constraint is violated. As far as we know, this inequality in the even sector does not have any geometric meaning. This confirms the observation that only the odd sector of the matrix model operators has a large $p$ limit that matches the volume of moduli spaces of hyperbolic surfaces with cone points.

\subsubsection{Five defects in a sphere}
We now extend the previous check to the case of four defects in the sphere. We start again with the minimal string matrix integral result, since it was already computed, and take the $p\to\infty$ limit. 

The minimal string answer was computed using the matrix model in \cite{Tarnopolsky:2009ec} and in the odd sector is given by:
\begin{align}
\frac{V_{0,5,0}^p(\boldsymbol{\alpha})}{(2\pi^2)^2} &= \sum_{i=1}^5 \left( H(k_i-2)-3 (1+p^{-1})F(k_i-1)  \right) +2\sum_{i<j}F(k_i-1)F(k_j-1) \nonumber\\
&\qquad+ \frac{(p+1)(5p^2+5p+2)}{2p^3}\nonumber\\
&\qquad-\sum_{I,|I|=2} \left( H(k_I-1)-(1+p^{-1})F(k_I)+F(k_I)\sum_{\ell\notin I} F(k_\ell-1)\right) \nonumber\\
&\qquad + \sum_{I,|I|=3}H(k_{I}) +\frac{1}{2}\sum_{I,J ,|I|=|J|=2,\,  I \cap J=\emptyset} F(k_{I})F(k_{J})\ ,
\end{align}
where we remind the reader that $I$ labels a set out of the 5 arguments and $k_I=\sum_{i\in I}k_i$. We define the function
\begin{equation}
H(k) = \frac{1}{2} F(k)F(k+2)\ ,\qquad \lim_{p\to\infty} H(p-p\alpha) = \frac{1}{2} \theta(\alpha)^2\ .
\end{equation}
The equation for $V_{0,5,0}^{p}$ has two differences when compared with the expression in \cite{Tarnopolsky:2009ec}. The first is trivial, we picked a different normalization of the tachyon to match with the defects in the large $p$ limit \cite{Turiaci:2020fjj}. More importantly, there is a typo in, for example, the last term of equation (3.6) of \cite{Tarnopolsky:2009ec}: the sum should not be under unrestricted four integers but over two pairs of integers $I$ and $J$ with $|I|=|J|=2$ with empty intersection $I \cap J=\emptyset$ and without overcounting choices of the two pairs (in the equation above in the last term we instead sum over all pairs including repetitions hence the factor of $1/2$ in front).

We are now ready to take the large $p$ limit of the expression above, giving:
\begin{align}
 \frac{V_{0,5,0}^p(\boldsymbol{\alpha})}{(2\pi^2)^2} &=\frac{5}{2} + \frac{1}{2} \sum_i \alpha_i^4 + 2 \sum_{i<j} \alpha_i^2\alpha_j^2-3\sum_i \alpha_i^2\nonumber\\
 &\qquad - \frac{1}{2} \sum_{I,|I|=2} \theta_I(\boldsymbol{\alpha})^2+\sum_{I,|I|=2}  \theta_I(\boldsymbol{\alpha})-\sum_{I,|I|=2}\sum_{\ell\neq i,j}\alpha_\ell^2 \,  \theta_I(\boldsymbol{\alpha})\nonumber\\
 &\qquad +\frac{1}{2}\sum_{I,J ,|I|=|J|=2,\,  I \cap J=\emptyset} \theta_I(\boldsymbol{\alpha})\theta_J(\boldsymbol{\alpha})+\frac{1}{2} \sum_{I,|I|=3}  \theta_I(\boldsymbol{\alpha})^2\ .
\end{align}
Comparing this expression with \eqref{eq:V05 cases} we conclude once again the matrix model answer matches (recovered taking $p\to\infty$) matches the WP volume defined in section \ref{sec:WPform}: 
\be 
    \lim_{p\to\infty} V_{0,5,0}^p(\boldsymbol{\alpha}) =V_{0,5,0}(\boldsymbol{\alpha}),
\ee
since the left hand side is (by definition) equal to the matrix integral proposed in section \ref{sec:mmvcp}.

Like the case with four defects, this match was verified in \cite{Artemev:2022hvu} only for sharp defects. In this case the only term is the first line of the equation above, which is the analytic continuation of the volume with geodesic boundaries. This is implied by the argument in \cite{Maxfield:2020ale}, but the validity of the rest of the terms here is a non-trivial check of the conjecture put forward here.

Partial success in reproducing the minimal string expression with five insertions from the continuum limit integrating the product of the mimimal model and Liouville five-point function over moduli space was presented in \cite{Artemev:2022rng}.

We will not write down explicit expressions but we have checked that the analogous formula in the even sector of the minimal string sphere five point function does not reproduce the full answer when defects are blunt (even though it matches the sharp regime).

\section{Conclusion}\label{sec:con}

We conclude with some comments and open questions raised by our work.

\textit{Derivation of WP form.} We found the Weil-Petersson measure on the volume of moduli space of hyperbolic surfaces with conical deficits in equation \eqref{eq:Weil-Petersson form with boundary classes}. We found it by solving a set of consistency conditions it should satisfy. It would be interesting to derive it from first principles, either from defining the moduli space in more detail in the presence of blunt defects, or from the JT gravity path integral perspective.

\textit{Proof of the recursion.} Another open question is to prove that the volumes computed using the measure \eqref{eq:Weil-Petersson form with boundary classes} satisfy the matrix integral topological recursion coming from the spectral curved derived from \eqref{eq:DOSGDJT}. For example, we do not know currently how to adapt Mirzakhani derivation in the presence of blunt defects.

\textit{Other ranges.} We have treated only the case of conical deficits. One may wonder whether these formulas have a further extension to defects outside of the range $0\leq \alpha\leq 1$ such as conical excesses. There are no corresponding operators in the minimal string and thus there is no clear expectation whether this should be possible. 

\textit{Dilaton gravity.} What is the right identification between the defect parameters and the dilaton potential? It was shown in \cite{Stanford:2022fdt} that for defects with very small deficit angles, the identification should be 
\begin{equation}\label{eq:WvsU}
U(\phi) = 2\phi + \sum_{i=1}^s (1-\alpha_i) \lambda_i \, \e^{-2\pi(1-\alpha_i)\phi}\ , \qquad 1-\alpha\ll 1\ .
\end{equation}
This follows also from the fact that the dilaton equation \eqref{eq:dilaton equation} requires a derivative which gets rid of the factor $1-\alpha_i$.
With this result, the defect generating function $W(y)$ is identified not with the dilaton potential as thought in \cite{Maxfield:2020ale,Witten:2020wvy} but with the prepotential instead,  $W(y) \sim \int^y \d \phi\, (U(\phi)-2\phi)$. An interesting physical question is whether the sum over the gas of defects (where each geometry is singular) is reproduced by a smooth geometry solving the new equations of motion with the modified dilaton potential $U(\phi)$. Evidence for this was provided in \cite{Turiaci:2020fjj} using \eqref{eq:WvsU} (even though it was not justified in that reference): the density of states derived from \eqref{eq:DOSGDJT} matches the semiclassical calculation of the gravity path integral in the modified smooth geometry derived from $U(\phi)$. 

\textit{Dilaton gravity and the minimal string.} We found that only the odd sector of the minimal string has a relation to hyperbolic geometry. It would be interesting to have a worldsheet understanding of this. Conversely it would be interesting to explore whether the even sector has any geometric meaning as well.

\paragraph{Acknowledgements} We thank J. Kruthoff, A. Levine, D. Stanford and E. Witten for discussions. We thank A. Artemev for pointing out some typos in a previous version. LE is supported by the grant DE-SC0009988 from the U.S. Department of Energy. GJT is supported by the Institute for Advanced Study and the National Science Foundation under Grant No. PHY-2207584, and by the Sivian Fund.

\appendix

\section{Relations of cohomology classes on \texorpdfstring{$\bM_{g,n}$}{Mg,n}} \label{app:relation cohomology}
In this appendix, we collect a number of useful identities between the cohomology classes $\kappa_1$, $\psi_1,\dots,\psi_n$ and $\delta_{h,I}$ on $\bM_{g,n}$. The only generic relation that these classes satisfy is
\be 
\delta_{h,I}=\delta_{g-h,I^c}\ .
\ee
\subsection{Relations for \texorpdfstring{$g=0$}{g=0}}
The relations at genus 0 play an important role. We have \cite{Arbarello_Cohomology}
\begin{subequations}
\begin{align}
    \psi_i&=\sum_{j,\, k \not \in I,\, i \in I} \delta_{0,I}\ , \label{eq:psi class identity M0n}\\
    \kappa_1&=\sum_{i,\, j \not \in I} (|I|-1) \delta_{0,I}\ . \label{eq:kappa class identity M0n}
\end{align}
\end{subequations}
Here $i,$ $j$ and $k$ are all assumed to be distinct, but arbitrary. These relations are complete and the rank of $\H^2(\bM_{0,n},\RR)$ is as a consequence $2^{n-1}-\binom{n}{2}-1$.
\subsection{Relations under forgetful pullback and pushforward}
Consider the forgetful morphism
\be 
\pi: \bM_{g,n+1} \longrightarrow \bM_{g,n}\ .
\ee
We can then consider the pullback of any of the classes above. This gives
\begin{subequations} \label{eq:pullback classes forgetful morphism}%
\begin{align}
    \pi^*(\kappa_1)&=\kappa_1-\psi_{n+1}\ , \\
    \pi^*(\psi_i)&=\psi_i-\delta_{0,\{i,n+1\}}\ , \\
    \pi^*(\delta_{h,I})&=\delta_{h,I}+\delta_{h,I+1}\ .
\end{align}
\end{subequations}
We can also consider the pushforward (fiberwise integration) of these classes. This yields degree zero classes, i.e.\ numbers
\begin{subequations}
\begin{align}
    \pi_*(\kappa_1)&=2g-2+n\ , \\
    \pi_*(\psi_i)&=1\ , \quad i=1,\dots,n\ , \\
    \pi_*(\psi_{n+1})&=2g-2+n\ , \\
    \pi_*(\delta_{h,I})&=\begin{cases}
    1 \ , \qquad & h=0\ \text{and}\ n+1 \in I \ \text{and}\ |I|=2\ , \\
    0 \ , \qquad &\text{otherwise}\ .
    \end{cases} 
\end{align}
\end{subequations}
\subsection{Relations under pullback to separating divisors}
A separating divisor $\mathscr{D}_{h,I}$ is isomorphic to $\bM_{h,|I|+1} \times \bM_{g-h,n-|I|+1}$. Correspondingly we get pullbacks $(\xi_{h,I}^\text{L})^*$ and $(\xi_{h,I}^\text{R})^*$ in cohomology to the two factors. It suffices to discuss $(\xi_{h,I}^\text{L})^*$ because $(\xi_{h,I}^\text{R})^*$ can be obtained from it by replacing $(h,I)$ with $(g-h,I^c)$. 
By renumbering indices we can also assume without loss of generality that $I=\{1,\dots,m\}$, so that
\be 
(\xi^\text{L}_{h,m})^*: \H^2(\bM_{g,n}) \longrightarrow \H^2(\bM_{h,m+1})\ .
\ee
We have then
\begin{subequations}
\begin{align}
    (\xi_{h,m}^\text{L})^*(\psi_i)&=\begin{cases}
     \psi_i \ , \qquad & i\le m\ , \\
     0 \ , \qquad & i>m\ ,
    \end{cases} \\
    (\xi_{h,m}^\text{L})^*(\kappa_1)&=\kappa_1\ .
\end{align}
\end{subequations}
The pullback of the class itself measures the self-intersection of the divisor $\mathscr{D}_{h,I}$. We have
\begin{align}
    (\xi_{h,m}^\text{L})^*(\delta_{h,\{1,\dots,m\}})=\begin{cases}
    -\psi_{m+1}+\delta_{2h-g,\{1,\dots,m+1\}}\ , \ & m=n\ \text{and}\  \frac{g}{2}\le h <g\ , \\
    -\psi_{m+1}\ , \ &\text{otherwise}\ .
    \end{cases} \label{eq:self intersection boundary class}
\end{align}
Assuming that $(h',J)\ne (h,\{1,\dots,m\})$ we have in the other cases
\begin{align}
    (\xi_{h,m}^\text{L})^*(\delta_{h',J})&=\begin{cases}
        \delta_{h',J}\ , \ &h' \le h\ \text{and}\ J \subset \{1,\dots,m\}\ , \\
        \delta_{h+h'-g,J \setminus \{m+2,\dots,n\}}\ , \ & h' \ge g-h \ \text{and}\ \{m+1,\dots,n\} \subset I\ , \\
        0 \ , \ & \text{otherwise}\ .
    \end{cases}
\end{align}
In the special case with $g$ even and $n=0$, the separating divisor $\mathscr{D}_{g/2,\emptyset}$ is isomorphic to $(\bM_{\frac{g}{2}}\times \bM_{\frac{g}{2}})/\ZZ_2$ where the $\ZZ_2$ symmetry exchanges the two copies of $\bM_{\frac{g}{2}}$. In this case we just get a single map $\xi_{h,m}^*=(\xi_{h,m}^\text{L})^*+(\xi_{h,m}^\text{R})^*$. The previous formulas are valid, provided that we insert a factor of 2 because the map $\bM_{\frac{g}{2}}\times \bM_{\frac{g}{2}} \longrightarrow \mathscr{D}_{g/2,\emptyset}$ has degree $2$.

\section{Some intersection numbers} \label{app:intersection numbers}
\subsection{Intersection numbers on \texorpdfstring{$\bM_{0,5}$}{M0,5}}
Let us provide a list of intersection numbers on $\bM_{0,5}$ that are needed to compute the Weil-Petersson volume. We assume that $|I|=2$ for the boundary class $\delta_{0,I}$ because $\delta_{0,I}=\delta_{0,I^c}$ which gives the case with $I=3$.
\begin{subequations}
\begin{align}
    \langle \kappa_1^2 \rangle &= 5\ , \\
    \langle \kappa_1 \psi_i \rangle &= 3\ , \\
    \langle \psi_i \psi_j \rangle &=\begin{cases}
    1 \ , i=j \, \\
    2 \ , i \ne j\ , 
    \end{cases} \\
    \langle \kappa_1 \delta_{0,I} \rangle &=1\ , \\
    \langle \psi_i \delta_{0,I} \rangle&= \begin{cases}
    1 \ , \quad i \not\in I\ , \\
    0 \ , \quad \text{otherwise}\ , 
    \end{cases} \\
    \langle \delta_{0,I} \delta_{0,J} \rangle &=\begin{cases}
    -1 \ , \quad I=J\ , \\
    1  \ , \quad I \cap J = \emptyset\ , \\
    0 \ , \quad \text{otherwise}\ .
    \end{cases}
\end{align}
\end{subequations}

\subsection{Intersection numbers on \texorpdfstring{$\bM_{1,2}$}{M1,2}}
The nonzero intersection numbers of $\kappa_1$, $\psi_i$ and $\delta_{0,\{1,2\}}$ are
\begin{subequations}
\begin{align}
    \langle \kappa_1^2 \rangle&=\frac{1}{8}\ , \\
    \langle \kappa_1 \psi_i \rangle&=\frac{1}{12}\ , \\
    \langle \psi_i\psi_j \rangle &=\frac{1}{24}\ , \\
    \langle \kappa_1 \delta_{0,\{1,2\}} \rangle &=\frac{1}{24}\ , \\
    \langle \delta_{0,\{1,2\}}^2 \rangle &=-\frac{1}{24}\ .   
\end{align}
\end{subequations}
\bibliographystyle{JHEP}
\bibliography{bib}
\end{document}